\documentclass[twocolumn,12pt]{iopart}

\usepackage{graphicx}

\usepackage[usenames]{color}

\newcommand{\beq}{\begin{equation}}
\newcommand{\eeq}{\end{equation}} 
\newcommand{\beqa}{\begin{eqnarray}}
\newcommand{\eeqa}{\end{eqnarray}} 
\begin{document}

\title{Statics and dynamics of a binary dipolar Bose-Einstein condensate soliton}

\author{ S. K. Adhikari\footnote{adhikari@ift.unesp.br; URL: http://www.ift.unesp.br/users/adhikari}
  and Luis E. Young-S.\footnote{lyoung@ift.unesp.br}
 }
\address{
Instituto de F\'{\i}sica Te\'orica, UNESP - Universidade Estadual Paulista, \\ 01.140-070 S\~ao Paulo, S\~ao Paulo, Brazil
} 

\begin{abstract}

We study the statics and dynamics of a binary dipolar Bose-Einstein 
condensate soliton for repulsive inter- and intraspecies contact interactions 
with the two components subject to different spatial symmetries $-$ distinct 
quasi-one-dimensional and quasi-two-dimensional shapes $-$ using numerical 
solution and variational approximation of a three-dimensional mean-field 
model. The results are illustrated with realistic values of parameters in the 
binary $^{164}$Dy-$^{168}$Er mixture. The possibility of forming robust 
dipolar solitons of very large number of atoms make them of great 
experimental interest. The existence of the solitons is illustrated in terms 
of stability phase diagrams. Exotic shapes of these solitons are illustrated 
in isodensity plots. The variational results for statics (size and chemical 
potential) and dynamics (small oscillation) of the binary soliton compare 
well with the numerical results. A way of preparing and studying these 
solitons in laboratory is suggested.

\end{abstract}


\maketitle

\section{Introduction}

A bright soliton is a self-reinforcing  solitary wave that 
travels at constant speed maintaining its shape, due to a cancellation of  dispersive effect and nonlinear attraction. 
Matter-wave soliton and quasi-one-dimensional (quasi-1D) soliton train were created and investigated 
experimentally
in 
Bose-Einstein condensate (BEC) of $^7$Li \cite{1, 2} and $^{85}$Rb atoms 
\cite{3}. These quasi-1D
solitons appear for attractive contact interaction 
in a  
axially-free BEC under radial harmonic trap \cite{4}.

The observation of  BECs of $^{164}$Dy \cite{dy,ExpDy}, $^{168}$Er \cite{ExpEr} and 
  $^{52}$Cr \cite{crrev,cr,saddle,ExpCr,52Cr} atoms
with large magnetic dipole
moments has opened new directions of research in the study of BEC solitons.
Polar molecules with much larger electric dipole moments are also being considered for BEC experiments \cite{polar}.
In addition to the conventional quasi-1D solitons \cite{1D}
of nondipolar BEC,
one can have quasi-two-dimensional (quasi-2D) solitons in dipolar BEC \cite{2D}. 
More interestingly, one can have dipolar BEC solitons for fully repulsive 
contact interaction \cite{1D}. Moreover, solitons in dipolar BEC remain stable when the 
harmonic trap(s) is(are) replaced by periodic optical-lattice trap(s)
 in quasi-1D \cite{ol1D} and quasi-2D \cite{ol2D} configurations.
Because of these interesting possibilities in dipolar BEC,
we study here the formation of solitons in a binary dipolar BEC.
Because of the complexity in dealing with the inter- and intraspecies dipolar interactions, there have been only a few studies of the binary dipolar 
mixture \cite{mfb,mfb2}. 
We consider the numerical solution and variational approximation 
of a three-dimensional (3D) mean-field model in our study of binary
dipolar BEC soliton where the atoms are polarized along $z$ axis.

There have been recent experimental \cite{ex} and theoretical \cite{th} studies in binary BECs employing
distinct trapping symmetry on each component. For example, the first component of the binary BEC could have a quasi-1D shape and the second component a quasi-2D shape, or one can have  a quasi-2D-quasi-2D binary mixture with the first component lying in the $x-y$ plane and the second component in the 
$y-z$ plane. Hence, we will also consider distinct spatial symmetry of the 
two components in a binary dipolar BEC soliton, thus leading to exotic density profiles of the mixture.  
Among the 
distinct spatial symmetries of the binary dipolar BEC soliton, we consider
(a) both components in quasi-1D shape along $z$ axis, (b) one component in quasi-2D shape in $y-z$ plane and the other  in quasi-1D shape along $z$ axis,
(c) both components in quasi-2D shape in $y-z$ plane, (d) one component in quasi-2D shape in $y-z$ plane and the other  in quasi-2D  shape in $x-z$ plane, and finally, 
(e)  one component in quasi-2D shape in $x-y$ plane and the other  in quasi-1D shape along $y$ axis. All these possibilities are realized for repulsive 
inter- and intraspecies contact interactions except the last one where we need 
attractive interspecies contact interaction for stability. 
This creates
a new scenario for robust solitons of very large numer of
atoms stabilized by short-range repulsion and long-range
inter- and intraspecies dipolar attraction.

We illustrate our findings using realistic parameters in the $^{164}$Dy-$^{168}$Er mixture. The stability of binary dipolar solitons is illustrated in phase diagrams involving critical number of atoms and interaction strengths. The profiles of the binary solitons are displayed in isodensity plots of the two 
components.  The variational approximation to the sizes and chemical potentials of the two components  is compared with the numerical solution of the mean-field 
model. The numerical study of breathing oscillation of the stable dipolar binary BEC soliton  is found to be in reasonable 
agreement with a time-dependent variational model calculation.

In section \ref{II}  the mean-field model for the binary    
dipolar BEC soliton  is developed.
A time-dependent, analytic,  Euler-Lagrange Gaussian variational approximation 
of the model is also presented. 
The results of numerical calculation are shown in section \ref{III}.  
Finally, in section \ref{IV} we present a brief summary of our findings.

\section{Mean-field model for a binary dipolar BEC soliton}

\label{II}

We consider a binary dipolar BEC soliton, interacting via  inter- and 
intraspecies interactions, with the 
mass, number of atoms, magnetic  dipole moment, and scattering length for the two species $ i=1,2,$
denoted by $m_i, N_i, 
\widehat \mu_i, a_i,$ respectively.   The inter- ($V_{12}$)
and intraspecies ($V_i$)
interactions 
for two atoms at positions $\bf r$ and $\bf r'$ are taken as
\begin{eqnarray}\label{intrapot} 
V_{12}({\bf R})= \frac{\mu_0\widehat \mu_1 \widehat \mu_2}{4\pi}\frac{1-3\cos^2 \theta}{|{\bf R}|^3}+
\frac{2\pi \hbar^2 a_{12}}{m_R}\delta({\bf R}), \\ \label{interpot} 
V_i({\bf R})= 
\frac{\mu_0\widehat \mu_i^2}{4\pi}\frac{1-3\cos^2 \theta}{|{\bf R}|^3}+\frac{4\pi 
\hbar^2 a_i}{m_i}\delta({\bf R }),
     \end{eqnarray}
where $\bf R = r-r',$ $\mu_0$ is the permeability of free space, 
$\theta$ is the angle made by the vector ${\bf R}$ with the polarization 
$z$ direction,  $a_{12}$ is the intraspecies scattering length
and $m_R=m_1m_2/(m_1+m_2)$ is the reduced mass of the two species of 
atoms. With these interactions, the coupled Gross-Pitaevskii (GP) 
equations for the binary dipolar BEC can be written as \cite{mfb2}
\beqa && \,
{\mbox i} \hbar \frac{\partial \phi_1({\bf r},t)}{\partial t}=
{\Big [}  -\frac{\hbar^2}{2m_1}\nabla^2+
\frac{1}{2}m_1\omega_1^2 (\nu_1x^2+\gamma_1 y^2+\lambda_1z^2)
\nonumber
\\  && 
+ \frac{4\pi \hbar^2}{m_1}{a}_1 N_1 \vert \phi_1({\bf r},t)\vert^2
+\frac{2\pi \hbar^2}{m_R} {a}_{12} N_2 \vert \phi_2({\bf r},t)|^2
%\nonumber \\ &&
+ N_1 \frac{ \mu_0 \ {\widehat \mu}^2_1 }{4\pi}
\int V_{dd}({\mathbf R})\vert\phi_1({\mathbf r'},t)\vert^2 d{\mathbf r}'
\nonumber \\ &&
+ N_2 \frac{ \mu_0 \ {\widehat \mu}_1\widehat \mu_2 }{4\pi}
\int V_{dd}({\mathbf R})\vert\phi_2({\mathbf r'},t)\vert^2 d{\mathbf r}'
{\Big ]}  \phi_1({\bf r},t),
\label{eq1}
\eeqa
\beqa
\label{eq2}
&&{\mbox i} \hbar \frac{\partial \phi_2({\bf r},t)}{\partial t}=
{\Big [}  -\frac{\hbar^2}{2m_2}\nabla^2+
\frac{1}{2}m_2\omega_2^2 (\nu_2x^2+\gamma_2 y^2+\lambda_2z^2)
\nonumber\\ &&
+ \frac{4\pi \hbar^2}{m_2}{a}_2 N_2 \vert \phi_2({\bf r},t) \vert^2
+\frac{2\pi \hbar^2}{m_R} {a}_{12} N_1 \vert \phi_1({\bf r},t) \vert^2
%\nonumber 
%\\ && 
+ N_2 \frac{ \mu_0 \ {\widehat \mu}^2_2 }{4\pi}
\int V_{dd}({\mathbf R})\vert\phi_2({\mathbf r'},t)\vert^2 d{\mathbf r}' 
\nonumber \\ &&
+ N_1 \frac{ \mu_0 \ {\widehat \mu}_1\widehat \mu_2 }{4\pi}
\int V_{dd}({\mathbf R})\vert\phi_1({\mathbf r'},t)\vert^2 d{\mathbf r}'
\Big] 
 \phi_2({\bf r},t),
\\&&
V_{dd}({\mathbf R})= 
\frac{1-3\cos^2\theta}{{\mathbf R}^3},  \quad
  \rho^2=x^2+y^2, \quad {\mbox i}=\sqrt{-1}.  
\eeqa
Here $\omega_i$ are the frequencies of the traps and  $\nu_i,\gamma_i,$ and
$\lambda_i$ are trap anisotropy parameters. 

To compare the dipolar and contact interactions, the intra- and interspecies 
dipolar interactions  are  expressed in terms of the dipolar lengths
$a_{dd}^{(i)}$ and $a_{dd}^{(12)}$, defined by 
\begin{eqnarray}  a_{dd}^{(i)}=
\frac{\mu_0\widehat \mu_i^2m_i}{12\pi \hbar ^2    },\quad
a_{dd}^{(12)}=
\frac{\mu_0\widehat \mu_1\widehat \mu_2m_R}{6\pi \hbar ^2    }.
\end{eqnarray}
We express the strengths of the dipolar 
interactions  by these lengths
and transform  (\ref{eq1}) and (\ref{eq2}) 
into the following dimensionless form  \cite{mfb2}
\beqa
&& \,
{\mbox i} \frac{\partial \phi_1({\bf r},t)}{\partial t}=
{\Big [}  -\frac{\nabla^2}{2 }
%+ v_1[\cos(2x)+\cos(2y)]
+
\frac{1}{2} (\nu_1x^2+\gamma_1 y^2+\lambda_1z^2)
\nonumber \\  &&  \,
+ g_1 \vert \phi_1 \vert^2
+ g_{12} \vert \phi_2 \vert^2
+ g_{dd}^{(1)}
\int V_{dd}({\mathbf R})\vert\phi_1({\mathbf r'},t)
\vert^2 d{\mathbf r}' 
\nonumber \\  &&  \,
+ g_{dd}^{(12)}
\int V_{dd}({\mathbf R})\vert\phi_2({\mathbf r'},t)
\vert^2 d{\mathbf r}' 
{\Big ]}  \phi_1({\bf r},t),
\label{eq3}
\eeqa
\beqa
&& \,
{\mbox i} \frac{\partial \phi_2({\bf r},t)}{\partial t}={\Big [}  
-m_{12} \frac{\nabla^2}{2}
+
\frac{1}{2}m_\omega (\nu_2x^2+\gamma_2 y^2+\lambda_2z^2)
\nonumber \\  &&  \,
+ g_2 \vert \phi_2 \vert^2 
+ g_{21} \vert \phi_1 \vert^2 
+ g_{dd}^{(2)}
\int V_{dd}({\mathbf R})\vert\phi_2({\mathbf r'},t)
\vert^2 d{\mathbf r}'
\nonumber \\ && \,
+ g_{dd}^{(21)}
\int V_{dd}({\mathbf R})\vert\phi_1({\mathbf r'},t)
\vert^2 d{\mathbf r}'  
{\Big ]}  \phi_2({\bf r},t),
\label{eq4}
\eeqa
where
%\begin{align}&
 $m_\omega=\omega_2^2/(m_{12}\omega_1^2),$
$m_{12}={m_1}/{m_2},$
$g_1=4\pi a_1 N_1,$
$g_2= 4\pi a_2 N_2 m_{12},$
$g_{12}={2\pi m_1} a_{12} N_2/m_R,$
$g_{21}={2\pi m_1} a_{12} N_1/m_R,$
$g_{dd}^{(2)}= 3N_2 a_{dd}^{(2)}m_{12},$
$g_{dd}^{(1)}= 3N_1 a_{dd}^{(1)},$
$g_{dd}^{(12)}= 3N_2 a_{dd}^{(12)}m_1/2m_R,$
$g_{dd}^{(21)}= 3N_1 a_{dd}^{(12)}m_{1}/2m_R.$
In  (\ref{eq3}) and (\ref{eq4}), length is expressed in units of 
oscillator length  $l_0=\sqrt{\hbar/(m_1\omega_1)}$, 
energy in units of oscillator energy  $\hbar\omega_1$, density 
$|\phi_i|^2$ in units of $l_0^{-3}$, and time in units of $ 
t_0=1/\omega_1$.

Convenient analytic variational approximation to  (\ref{eq3}) and (\ref{eq4}) can be obtained with the following 
ansatz for the wave functions in case of axially symmetric traps with 
$\nu_i=\gamma_i=1$: 
\cite{17,Santos01,pg}
\beqa \phi_i({\bf r},t)=\frac{\pi^{-3/4}}{w_{\rho i}\sqrt{w_{z 
i}}}\exp\Big[-\frac{\rho^2} {2w_{\rho 
i}^2}-\frac{z^2}{2w_{zi}^2}+\mathrm{i}\alpha_i\rho^2+\mathrm{i}\beta_i 
z^2\Big] \eeqa
where $ {\bf r}=\{\vec \rho,z   \}, {\vec \rho}=\{x,y\}$, $w_{\rho i}$ and $w_{z i}$ are the widths and $\alpha_i$ and 
$\beta_i$ are additional 
variational parameters. The effective Lagrangian for the binary system is 
\cite{pg}
\beqa
\fl L=\int d{\bf r} \frac{1}{2}\Big[
\sum_i \Big\{ {\mathrm i}N_i(\phi_i\dot \phi_i^*- \phi_i^* \dot \phi_i)
+ N_ig_i|\phi_i({\bf r})|^4\Big\}
%\nonumber \\ &&
+N_1 [\rho^2+\lambda_1 z^2]|\phi_1({\bf r})|^2
\nonumber \\
\fl
+N_2 [\rho^2+\lambda_2 z^2]|\phi_2({\bf r})|^2m_\omega
%\nonumber \\&&
+ N_1|\nabla \phi_1({\bf r})|^2+m_{12} N_2|\nabla \phi_2({\bf r})|^2+ 2N_1g_{12} |\phi_1({\bf r})|^2   |\phi_2({\bf r})|^2
\Big]
\nonumber \\
\fl
+ \int \int d{\bf r}' d{\bf r}
 \Big[  \sum_i\frac{N_i}{2}g_{dd}^{(i)} V_{dd}({\bf R})|\phi_i({\bf r'})|^2   |\phi_i({\bf r})|^2 
%\nonumber \\
%&&
+ {N_1}g_{dd}^{(12)} V_{dd}({\bf R})|\phi_1({\bf r'})|^2   |\phi_2({\bf r})|^2 
\Big] 
, \\
%\end{align}
%\begin{align}
 \label{lag}
\fl = 
\sum_{i=1}^2 \frac{N_i}{2}(2w_{\rho i}^2\dot \alpha_i + w_{z i}^2\dot\beta_i )
%\nonumber \\ &&
+N_1\left[\frac{w_{\rho 1}^2}{2}+\frac{\lambda_1 w_{z1}^2}{4}  \right]
+m_\omega N_2\left[\frac{w_{\rho 2}^2}{2} +\frac{\lambda_2 w_{z2}^2}{4}  \right]
\nonumber \\ 
\fl +
\frac{N_1}{2}
\biggr[\frac{1}{w_{\rho 1}^2}+\frac{1}{2w_{z1}^2}
+4w_{\rho 1}^2\alpha_1^2+
2w_{z1}^2\beta_1^2 \biggr]
%\nonumber \\ &&
+\frac{N_2m_{12}}{2}\biggr[\frac{1}{w_{\rho 2}^2}+\frac{1}{2w_{z2}^2}
+ 4w_{\rho 2}^2\alpha_2^2+
2w_{z2}^2\beta_2^2 \biggr]
\nonumber \\
%\eeqa
%\beqa
\fl +\frac{N_1^2[ a_1-a_{dd}^{(1)}f(\kappa_1)]}
{\sqrt{2\pi}w_{\rho 1}^2w_{z 1}}+\frac{N_2^2 m_{12}
[a_2-a_{dd}^{(2)}f(\kappa_2)]
}
{\sqrt{2\pi}w_{\rho 2}^2w_{z 2}}
 +\frac{2m_1N_1N_2[a_{12}-a_{dd}^{(12)}f(\kappa_3)]}{\sqrt \pi m_R w_{\rho 3}^2w_{z 3}}, \quad 
     \\
f(\kappa)=\frac{1+2\kappa^2-3\kappa^2d(\kappa)}{1-\kappa^2}, \quad d(\kappa)=\frac{\mathrm{atan}(\sqrt{\kappa^2-1})}
{\sqrt{\kappa^2-1}}, \nonumber
\eeqa
where $\kappa_i=w_{\rho i}/w_{z i}, w_{\rho 3} =\sqrt{w_{\rho 1}^2+w_{\rho 2}^2}$ , $w_{z 3}
=\sqrt{ w_{z 1}^2+w_{z 2}^2}$.
% , and $C=4a_{12}m_1/(\sqrt \pi m_R)$. 
In these equations the overhead dot denotes time derivative.

The Euler-Lagrange variational equations for the widths for the effective Lagrangian (\ref{lag}), 
obtained in usual fashion \cite{pg}, 
can be written as 
\beqa\label{eq10}\fl
\ddot{w}_{\rho 1}=-\omega_{\rho 1}  + 
\frac{1}{w_{\rho 1}^3} +  \frac{N_1[2 a_1-a_{dd}^{(1)}g(\kappa_1)]
}{\sqrt{2\pi}
w_{\rho 1}^3w_{z1}} 
%\nonumber \\ &&
+
\frac{2m_1 N_2 w_{\rho 1}[2a_{12}-a_{dd}^{(12)}g(\kappa_3)] }{\sqrt \pi m_R   w_{\rho 3}^4  w_{z 3}  }
, \\ \label{eq11}
\fl \ddot{w}_{z1} = -\lambda_1\omega_{z1}+
\frac{1}{w_{z1}^3}+ 
\frac{2N_1[a_1-a_{dd}^{(1)}h(\kappa_1)]}{ \sqrt{2\pi} w_{\rho 1}^2w_{z1}^2} 
%\nonumber \\ &&
+\frac{{{4}}m_1 N_2w_{z1}[a_{12}-a_{dd}^{(12)}h(\kappa_3)]}{\sqrt \pi m_R w_{\rho 3}^2  w_{z 3}^3  },\\ 
\label{eq12}
\fl \ddot{w}_{\rho 2}=-m_\omega \omega_{\rho 2} 
+ 
\frac{m_{12}}{w_{\rho 2}^3} + \frac{N_2m_{12} [2a_2-
a_{dd}^{(2)} g(\kappa_2)]}{\sqrt{2\pi}w_{\rho 2}^3w_{z2}}
%\nonumber  \\&& 
+
\frac{2m_1 N_1 w_{\rho 2} [2a_{12}-a_{dd}^{(12)}g(\kappa_3)]
 }{\sqrt \pi m_R w_{\rho 3}^4  w_{z 3} }
,
 \\ \label{eq13}
\fl  \ddot{w}_{z2}  = -m_\omega\lambda_2\omega_{z2}
+\frac{m_{12}}{ w_{z2}^3} 
+\frac{2N_2m_{12} [a_2-
a_{dd}^{(2)}  h(\kappa_2)]}{\sqrt{2\pi}w_{\rho 2}^2w_{z2}^2}
%\nonumber \\ &&
+\frac{{{4}}m_1N_1  w_{z 2}[a_{12}-a_{dd}^{(12)}h(\kappa_3)] }{\sqrt \pi m_R
  w_{\rho 3}^2  w_{z 3}^3 },\\
g(\kappa)=  \frac{2-7\kappa^2-4\kappa^4+9\kappa^4d(\kappa)}{(1-\kappa^2)^2}, \\
h(\kappa)=  \frac{1+10\kappa^2-2\kappa^4-9\kappa^2d(\kappa)}{(1-\kappa^2)^2}. 
\eeqa 
The solution of the time-dependent 
equations (\ref{eq10}) $-$ (\ref{eq13}) gives the dynamics of the variational approximation.

The energy of the system is given by 
\beqa\label{energy}
 E=
N_1\left[\frac{w_{\rho 1}^2}{2}+\frac{\lambda_1 w_{z1}^2}{4}  \right]
+m_\omega N_2\left[\frac{w_{\rho 2}^2}{2} +\frac{\lambda_2 w_{z2}^2}{4}  \right]
%\nonumber \\&&
+
\frac{N_1}{2}
\biggr[\frac{1}{w_{\rho 1}^2}+\frac{1}{2w_{z1}^2} \biggr]
\nonumber \\
+\frac{N_2m_{12}}{2}\biggr[\frac{1}{w_{\rho 2}^2}+\frac{1}{2w_{z2}^2} \biggr]
+\frac{N_1^2[ a_1-a_{dd}^{(1)}f(\kappa_1)]}
{\sqrt{2\pi}w_{\rho 1}^2w_{z 1}}+\frac{N_2^2 m_{12}
[a_2-a_{dd}^{(2)}f(\kappa_2)]
}
{\sqrt{2\pi}w_{\rho 2}^2w_{z 2}}
\nonumber \\
+\frac{2m_1N_1N_2[a_{12}-a_{dd}^{(12)}f(\kappa_3)]}{\sqrt \pi m_Rw_{\rho 3}^2w_{z 3}
 }.
\eeqa
The energy is actually the stationary (time-independent) part of the 
Lagrangian (\ref{lag}).

If $\mu_i$  is the chemical potential with which the stationary wave function $\phi_i({\bf r},t)$ 
propagates in time, e.g.  $\phi_i({\bf r},t)\sim \exp(-i\mu_it )\phi_i({\bf r})$, 
then the variational estimate for $\mu_i  (\equiv \partial E/\partial N_i )$ is:
\beqa
&&\mu_1=\frac{\partial E}{\partial N_1}  =\frac{1}{2}
\biggr[\frac{1}{w_{\rho 1}^2}+\frac{1}{2w_{z1}^2}\biggr]
+\left[\frac{w_{\rho 1}^2}{2}+\frac{\lambda_1 w_{z1}^2}{4}  \right]
\nonumber 
\\&& +\frac{2N_1[a_1-a_{dd}^{(1)}f(\kappa_1)  ]}
{\sqrt{2\pi}w_{\rho 1}^2w_{z 1}}
+\frac{2m_1N_2[a_{12}-a_{dd}^{(12)}f(\kappa_3)]}{\sqrt \pi m_Rw_{\rho 3}^2w_{z 3}}
  ,\\
&&\mu_2=\frac{\partial E}{\partial N_2} 
=\frac{m_{12}}{2}\biggr[\frac{1}{w_{\rho 2}^2}+\frac{1}{2w_{z2}^2}
\biggr] 
+m_\omega \left[\frac{w_{\rho 2}^2}{2} +\frac{\lambda_2 w_{z2}^2}{4}  \right]
\nonumber \\&&+ \frac{2N_2 m_{12}
[a_2-a_{dd}^{(2)}f(\kappa_2)]
}
{\sqrt{2\pi}w_{\rho 2}^2w_{z 2}}
+\frac{2m_1N_1[a_{12}-a_{dd}^{(12)}f(\kappa_3)]}{\sqrt \pi m_Rw_{\rho 3}^2w_{z 3}
 }.
\eeqa
The widths of the stationary binary dipolar soliton 
can be 
obtained from the solution of (\ref{eq10}) $-$ (\ref{eq13}) 
setting the time derivatives of the widths 
 to zero.  This procedure is equivalent to a minimization of the energy (\ref{energy}), provided the 
stationary binary soliton is stable and corresponds to a energy minimum. In section \ref{III}, we will consider diverse types of anisotropic binary solitons. However, the Gaussian variational approximation above is applicable to only 
the conventional cigar-shaped axially-symmetric solitons with trap parameters
$\nu_i=\gamma_i=1$ and $\lambda_i=0$.

\section{Numerical Results}
\label{III}

We perform numerical calculation for the stability
and dynamics of the binary dipolar soliton using realistic values of 
atom numbers and interaction 
parameters 
in the $^{164}$Dy-$^{168}$Er mixture. The $^{164}$Dy and $^{168}$Er atoms 
have the largest magnetic moments of all the dipolar atoms used in BEC experiments. 
 The $^{164}$Dy atoms are labeled $i=1$ and  the $^{168}$Er atoms 
are labeled $i=2$.
  The $^{164}$Dy atoms have a large magnetic dipole moment $\widehat \mu_1 = 10\mu_B$
\cite{ExpDy}
with $\mu_B$ $(=9.27402 \times 10^{-24}$ Am$^2$) the Bohr magneton
corresponding to the dipolar length $a_{dd}^{(1)}\equiv \mu_0\widehat \mu_1^2 m_1/(12\pi
\hbar^2)\approx 132.7a_0$, with $a_0$ $ (=5.29\times 10^{-11}$ m) the Bohr
radius, $\mu_0=4\pi\times 10^{-7}$ N/A$^2,$ $\hbar= 1.05457\times
10^{-34}$ m$^2$kg/s,  1 amu $= 1.66054 \times 10 ^{-27}$ kg.   
For $^{168}$Er atoms, $\widehat \mu_2=7\mu_B$ \cite{ExpEr}, 
the dipolar length 
$a_{dd}^{(2)}\equiv \mu_0\widehat \mu_2^2 m_2/(12\pi
\hbar^2)
\approx 66.6a_0$, and the interspecies 
dipolar length
$a_{dd}^{(12)}\equiv
\mu_0\widehat \mu_1\widehat \mu_2m_R/(6\pi \hbar^2)
\approx 94.0a_0$.
Thus the dipolar interaction in $^{164}$Dy
atoms is more than eight times larger than that in $^{52}$Cr
atoms with a dipolar length  $a_{dd}\approx 15a_0$ \cite{cr,ExpCr}.  

 The contribution of the dipolar interaction is calculated in momentum 
space by Fourier transformation (FT) and the following convolution integral \cite{Santos01} 
\begin{eqnarray}
\int d{\bf r}V_{dd}({\bf r-r'})n({\bf r'})=\int \frac{d{\bf k}}{(2\pi)^3}e^{i{\bf k}\cdot {\bf r}}\tilde V_{dd}(\bf k)\tilde n({\bf k}),\\
\tilde V_{dd}({\bf k})=\frac{4\pi}{3}\left[\frac{3k_z^2}{{\bf k}^2}-1     \right],
\end{eqnarray}
with density $n({\bf r})=|\phi({\bf r})|^2$. The FT is defined by
\begin{eqnarray}
\tilde A({\bf k})=\int d{\bf r}A({\bf r})e^{i{\bf k}\cdot {\bf r}}, \\
A({\bf r})=\int \frac{d{\bf k}}{(2\pi)^3}\tilde A({\bf k})e^{-i{\bf k}\cdot {\bf r}}.
\end{eqnarray}
The FT $\tilde n({\bf k})$ of density and the inverse FT are 
 calculated numerically by a fast FT
 routine.  The whole procedure is performed in a 
3D Cartesian coordinate system irrespective of
the underlying trap symmetry. 
We solve  (\ref{eq3}) and (\ref{eq4})  
by the split-step 
Crank-Nicolson discretization scheme  
using a space step of $\sim 0.1 - 0.2$ 
and the time step $\sim 0.001 -0.003$ \cite{Santos01,CPC}.

We will be studying 
the binary solitons in different trap symmetries mostly for repulsive interspecies and intraspecies contact interactions. The attraction for the formation of the solitons  will be provided by the long-range dipolar interactions. These repulsive contact interactions will make the collapse more difficult and will generate robust binary dipolar solitons. This will allow us to consider binary solitons in a new scenario, which is not possible in a nondipolar BEC. However, that a net attraction for the formation of soliton is available  in the binary system, we take the 
dipolar length scales larger than the corresponding atomic scattering lengths:
$a_{dd}^{(i)}>a_i, a_{dd}^{(12)}>a_{12}$. If these conditions are satisfied 
the net nonlinear interaction turns to be attractive in an axially free set up 
as can be realized from 
 (\ref{lag}).
In the present study we take for the $^{164}$Dy atoms $a_1=120a_0$, and for the $^{168}$Er atoms $a_2=60a_0$.
The yet unknown interspecies scattering length $a_{12}$ is taken as a variable.  
Of these, the scattering length $a_1$ of  $^{164}$Dy atoms is close to the experimental value
  \cite{ExpDy}. The scattering lengths $a_2$ and $a_{12}$ can be controlled by independent magnetic \cite{fesh} and optical Feshbach resonance \cite{opfesh}
techniques.
  We consider the trap frequencies
$\omega_1=\omega_2=2\pi \times 61$ Hz, so that the length scale $l_0\equiv
\sqrt{\hbar/m_1\omega_1}=1$ $\mu$m,
% and time scale $t_0
%\equiv \omega_1^{-1}= 2.61$ ms 
and the constant $m_\omega= 1/m_{12}$
in (\ref{eq4}).

\begin{figure}[!t]

\begin{center}
\includegraphics[width=\linewidth,clip]{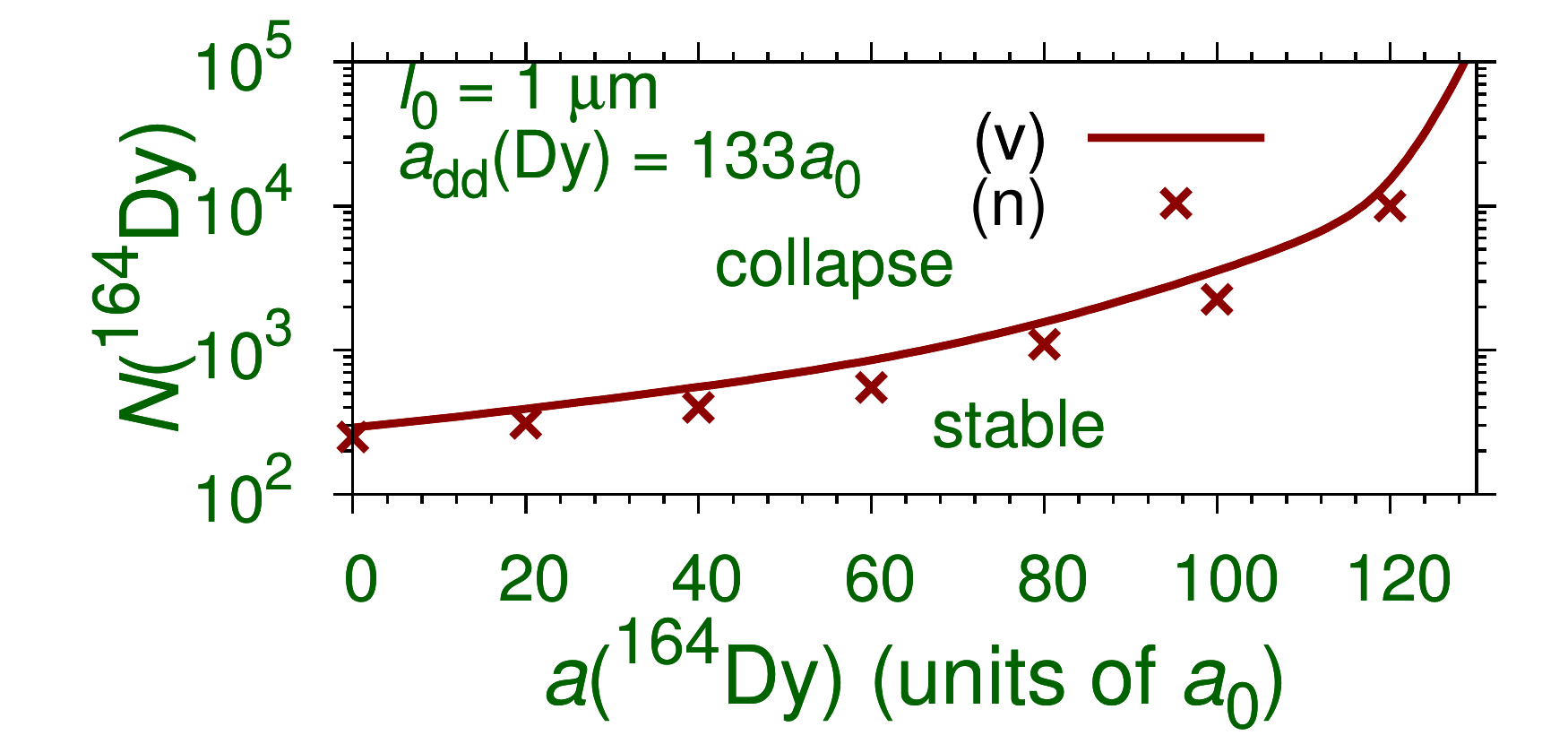}

\caption{   Stability phase diagram for a single-component 
$^{164}$Dy soliton free to move along the axial $z$ direction
for a harmonic trap $\rho^2/2$ in the $x-y$ plane
as obtained from numerical calculation (n) and variational approximation (v).
The oscillator length $l_0=1$ $\mu$m, $a_{dd}$(Dy)=$132.7a_0$.
}\label{fig1} \end{center}

\end{figure}

\begin{figure}[!t]

\begin{center}
\includegraphics[width=\linewidth]{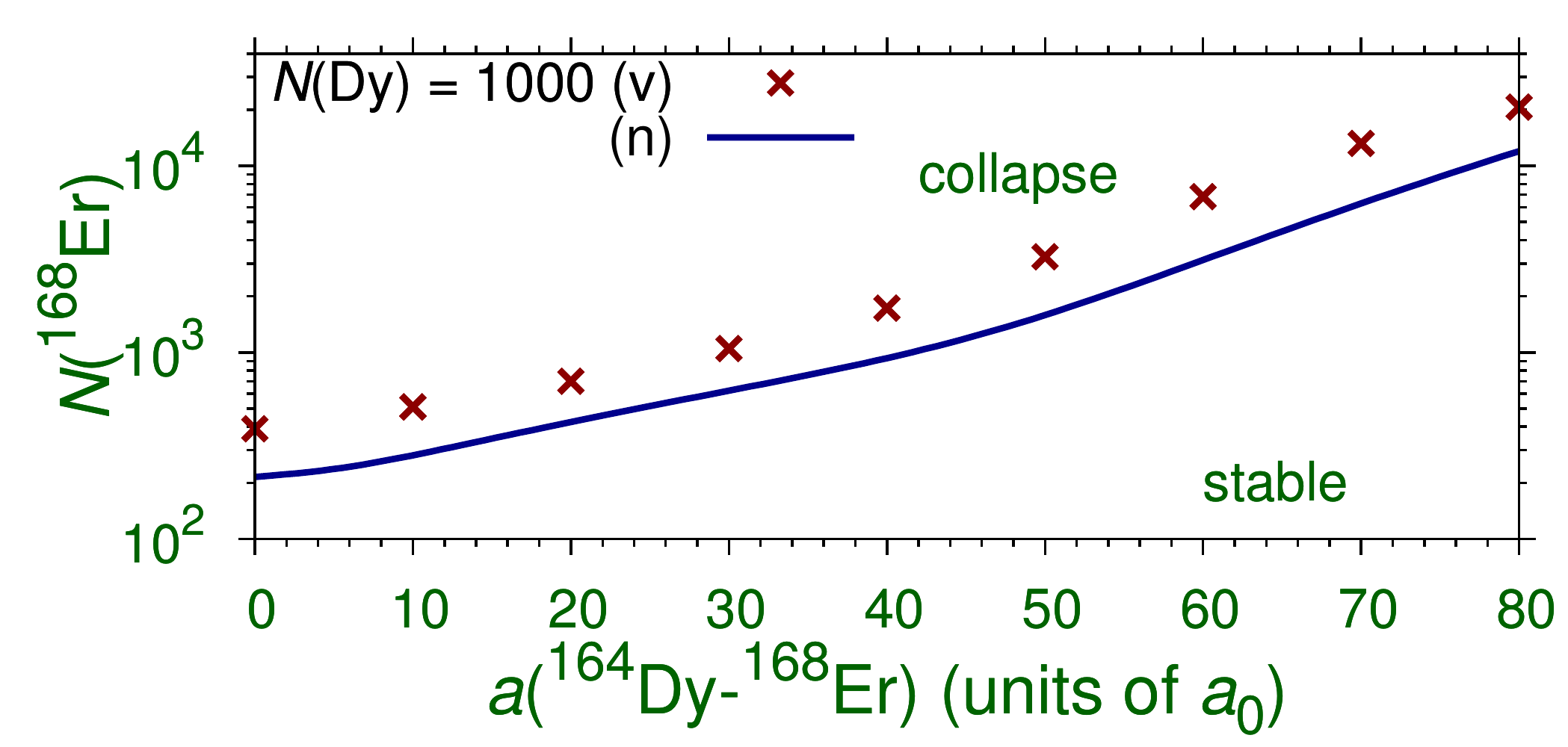}

\caption{ 
   Stability phase plot for a binary
$^{164}$Dy-$^{168}$Er soliton free to move along the axial $z$ direction
with the   axially-symmetric  trap $\rho^2/2$ in the $x-y$ plane on both components
as obtained from numerical calculation (n) and variational approximation (v).
The parameters employed are $a$(Dy)$=120a_0, a$(Er)$=60a_0, a_{dd}$(Dy)$= 
132.7a_0, a_{dd}$(Er)$=66.6a_0, a_{dd}$(Dy-Er)$=94a_0$,
$l_0= 1$ $\mu$m and $m_\omega=1/m_{12}$.
}\label{fig2} \end{center}

\end{figure}

%\begin{figure}[!t]

%\begin{center}
%\includegraphics[width=\linewidth]{fig3.eps}

%\caption{  Numerically obtained
%   stability phase plot for a binary
%$^{164}$Dy-$^{168}$Er soliton free to move along the axial $z$ direction
%with the  harmonic 1D trap $x^2/2$ on the first component ($^{164}$Dy) 
%and the  axially-symmetric  trap $\rho^2/2$ on the second component (Er). 
%All parameters are the same as in figure \ref{fig2}.
%}\label{fig3} \end{center}

%\end{figure}

\begin{figure}[!t]

\begin{center}
\includegraphics[width=\linewidth]{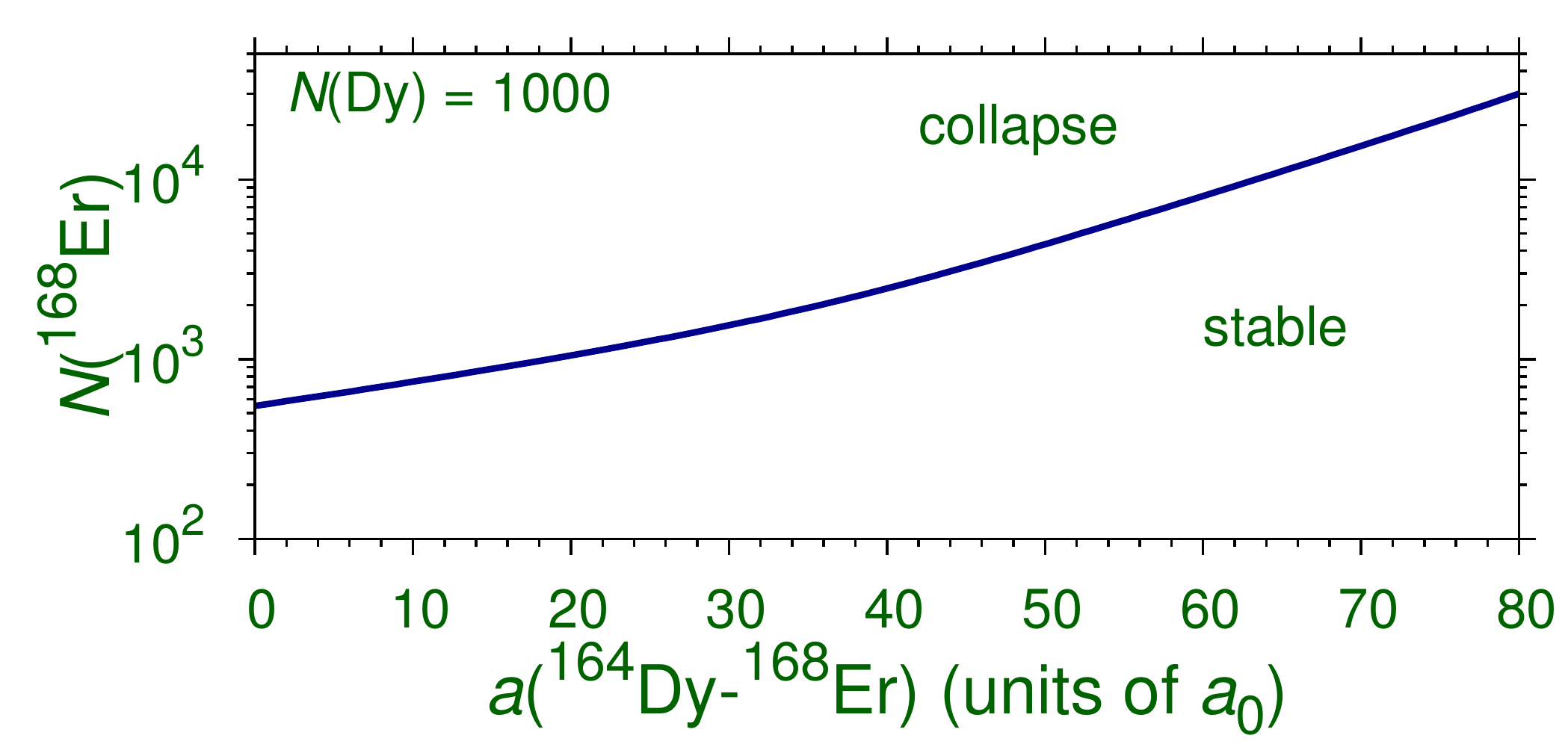}

\caption{ 
Numerically obtained   stability phase diagram for a binary quasi-2D
$^{164}$Dy-$^{168}$Er soliton free to move in the $y-z$ plane
with the  harmonic trap $x^2/2$ on both components. 
All parameters are the same as in figure \ref{fig2}.
}\label{fig4} \end{center}

\end{figure}

First we study the stability of a single-component  $^{164}$Dy soliton harmonically trapped in the $x-y$ plane 
by the 
potential  $\rho^2/2$  and free to move in the $z$ direction. The result is shown in figure \ref{fig1} where we
display the critical number of  atoms $N$(Dy) versus the atomic scattering length $a$(Dy) as obtained from
the Gaussian variational approximation and the complete numerical solution.   The variational energies are larger than the 
numerical energies and the variational states are loosely bound compared the numerical states. Consequently, the variational
states can accommodate more dipolar atoms in  a stable dipolar soliton as seen in figure \ref{fig1}.

Now we study the 
stability of a binary $^{164}$Dy-$^{168}$Er soliton free to move in the $z$ direction
while both components are harmonically trapped in the $x-y$ plane 
by the potential  $\rho^2/2$.  Both numerical and variational results are shown in figure \ref{fig2} for the number of  $^{164}$Dy atoms $N$(Dy) = 1000. In this figure we plot the critical number of  $^{168}$Er
atoms $N$(Er)  in the stable binary soliton versus the interatomic scattering length $a$(Dy-Er). As in the single-component case, the
variational critical number of   $^{168}$Er atoms in the binary soliton is larger than the numerical critical number of the same. 

In the above examples we considered only  the  soliton(s) with an axially-symmetric trap in the $x-y$ plane. In case of single-component dipolar soliton, the axial symmetry can be removed by taking only a harmonic or optical-lattice (OL) trap along the $x$  direction, thus generating an asymmetric two-dimensional soliton free to move in the 
$y-z$ plane \cite{2D}. 
In case of binary dipolar solitons the axial symmetry of the trap acting
on one or both components can be removed,  thus generating a new class of 
asymmetric  binary solitons.  One example of such asymmetric binary soliton is obtained by considering the harmonic 
trap $x^2/2$ on both components. 
In this case both component solitons  $^{164}$Dy and $^{168}$Er are clearly asymmetric.
  A stability phase diagram in this case is shown in figure \ref{fig4}
for 1000 $^{164}$Dy atoms, where  we plot the critical number $N$(Er)
of $^{168}$Er atoms in a stable binary soliton versus the interspecies scattering 
length $a$(Dy-Er).
In the following we will also consider few examples of binary dipolar solitons
with other types of traps. 
%Now we show the stability phase diagram of a quasi-2D binary soliton free to move in the $y-z$ %plane with the harmonic trap $x^2/2 $ on both components in figure \ref{fig4}. In this case both the solitons are clearly anisotropic due to the anisotropic trap $x^2/2$.
 
In figures \ref{fig2} and \ref{fig4} we have considered different types of
traps to achieve the binary dipolar soliton. As the number of  traps
on the binary soliton is reduced the solitons become
loosely bound thus accommodating more atoms. The increase of the number of 
traps makes the binary
soliton more compact and thus vulnerable to collapse due
to strong dipolar interaction. The binary soliton of figure \ref{fig2}
 is most compact with four 1D traps, that figure \ref{fig4} has
two 1D traps.
%, and that of figure \ref{fig4} is the least compact
%with only two 1D traps. 
Consequently, for the fixed number
$N$(Dy) = 1000 of $^{164}$Dy atoms, the critical number
of numerically obtained $^{168}$Er atoms is  smaller in
figure  \ref{fig2} compared to those in figure \ref{fig4}.

\begin{figure}[!t]

\begin{center}
\includegraphics[width=.49\linewidth]{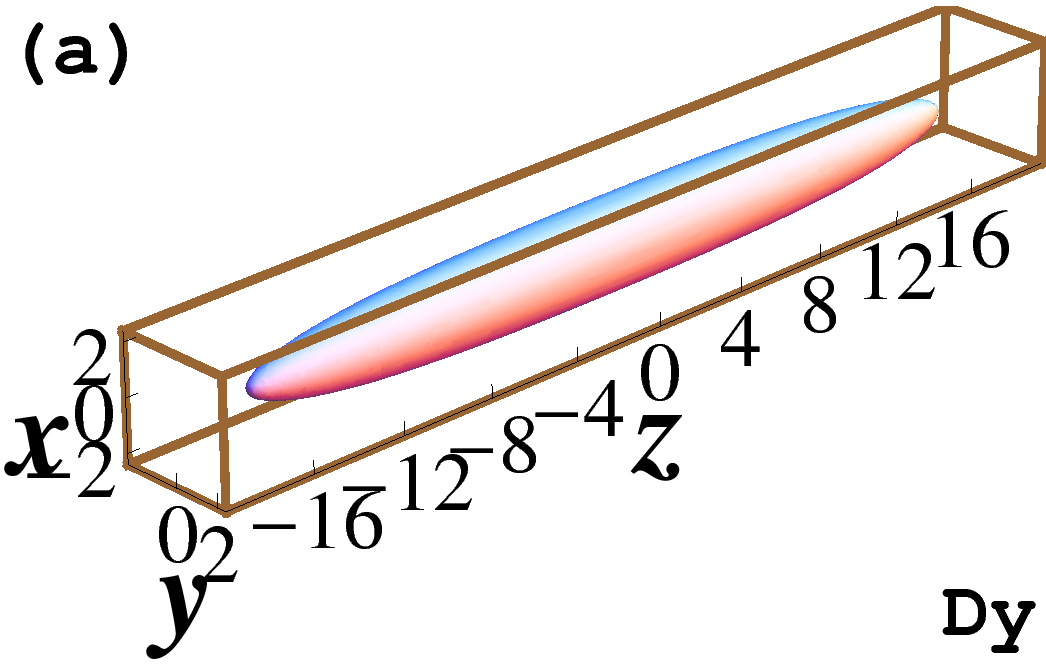}
\includegraphics[width=.49\linewidth]{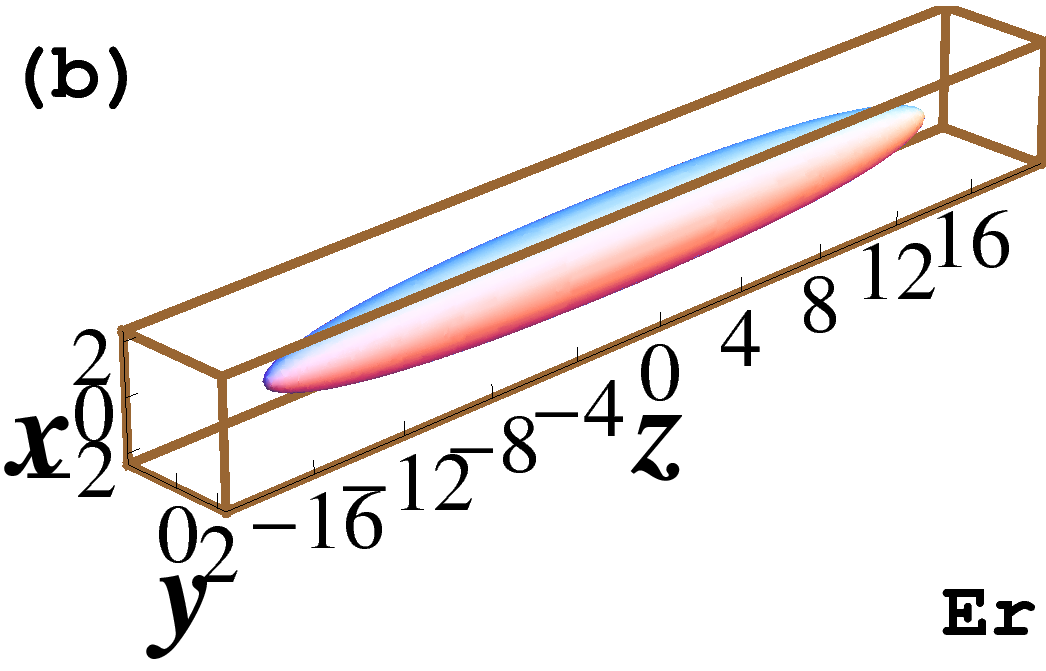}

\caption{  Isodensity plots of numerically obtained 
(a) $^{164}$Dy and (b) 
$^{168}$Er profiles 
 of a binary cigar-shaped 
$^{164}$Dy-$^{168}$Er soliton
 of 1000 atoms each, with the radial trap $\rho^2/2$ 
on both components, and 
free to move along the $z$ direction. All parameters are the same as 
in figure \ref{fig2} except the interspecies scattering length 
  $a$(Dy-Er) = $70a_0$. Lengths are in units of $l_0 (\equiv 1$ $\mu$m)  and density
in units of $l_0^{-3}$.  
The density on contour is 0.001.
}\label{fig5} \end{center}

\end{figure}

\begin{figure}[!t]

\begin{center}
\includegraphics[width=0.49\linewidth]{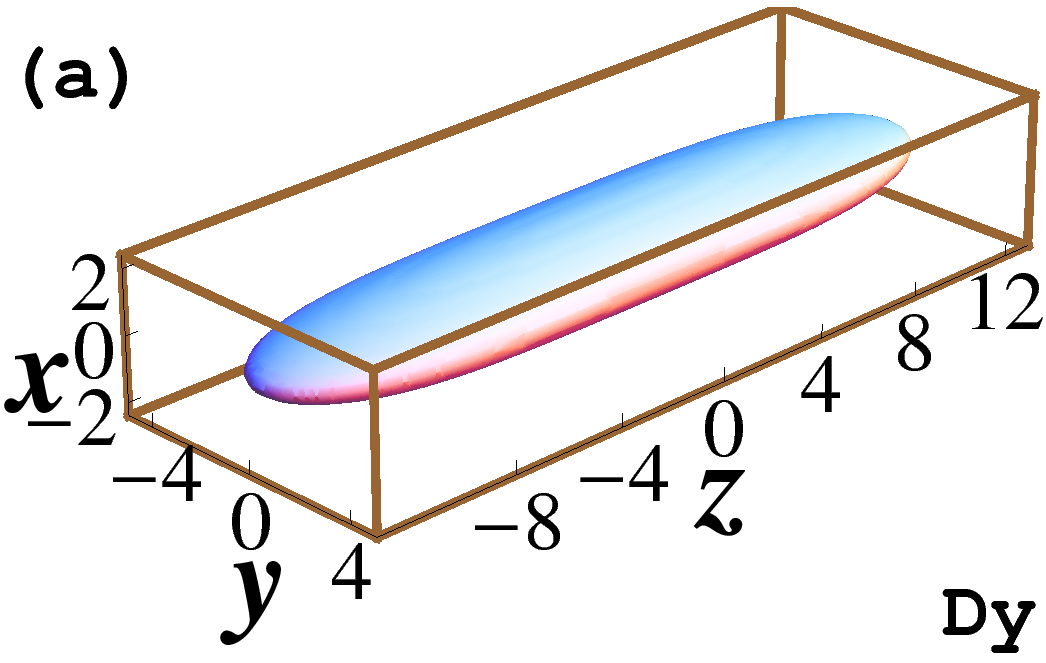}
\includegraphics[width=.49\linewidth]{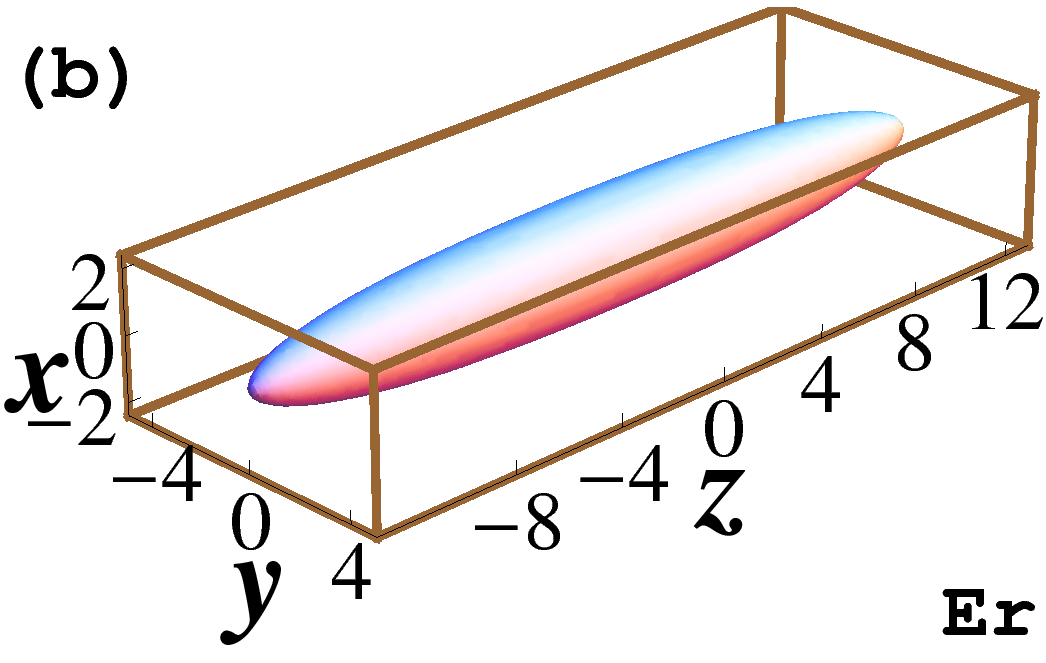}

\caption{ 
 Isodensity plots of numerically obtained 
(a) $^{164}$Dy and (b) 
$^{168}$Er profiles 
 of a binary 
$^{164}$Dy-$^{168}$Er soliton
 of 1000 atoms each and 
free to move along the $z$ direction. The first component 
(Dy), subject to a harmonic trap $x^2/2$, has a quasi-2D 
profile in $y-z$ plane whereas 
the second component (Er) with the  trap $\rho^2/2$ has a 
 quasi-1D profile along $z$ axis. 
All parameters are the same as 
in figure \ref{fig2} except the interspecies scattering length 
  $a$(Dy-Er) = $50a_0$. 
The density on contour is 0.001.
}\label{fig6} \end{center}
\end{figure}

The numerically obtained 
profiles of different types of binary dipolar solitons are next illustrated  by their isodensity contours. First we consider 
the axially-symmetric binary soliton with the axially-symmetric trap $\rho^2/2$ acting on both components
and free to move along the $z$ direction.  The corresponding isodensity 
profiles of the two components are shown in Figs. \ref{fig5} (a) and (b) for  1000 $^{164}$Dy and 1000 $^{168}$Er atoms, respectively. The solitons have axially-symmetric cigar shapes.  Next we consider an asymmetric binary dipolar soliton with the  
harmonic 1D 
trap $x^2/2$ on the first component (Dy) and the axially-symmetric trap  $\rho^2/2$ on the second component (Er). 
The corresponding isodensity 
profiles of the two components are shown in Figs. \ref{fig6} (a) and (b).

\begin{figure}%[!t]

\begin{center}
\includegraphics[width=.49\linewidth]{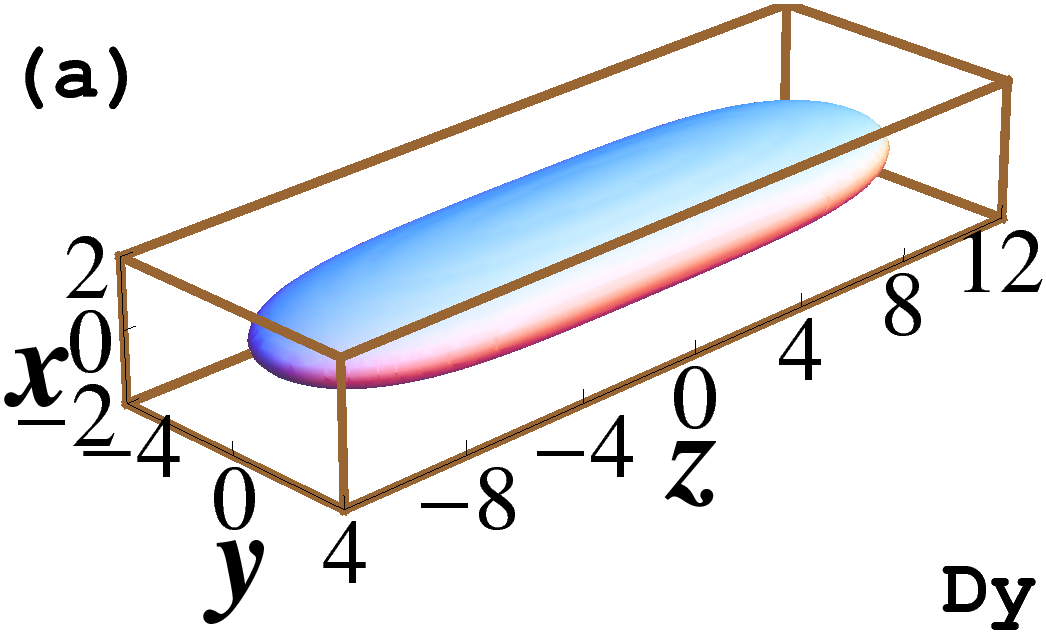}
\includegraphics[width=.49\linewidth]{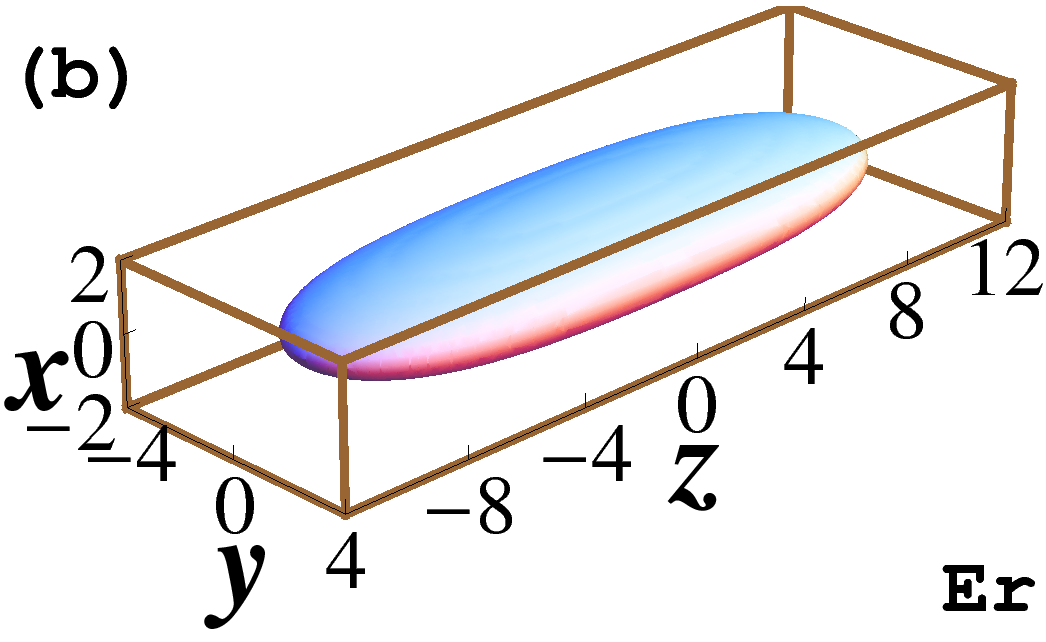}

\caption{  Isodensity plots of numerically obtained 
(a) $^{164}$Dy and (b) 
$^{168}$Er profiles of a binary 
$^{164}$Dy-$^{168}$Er soliton
of 1000 atoms each and 
free to move in the $y-z$ plane. 
Both components, subject 
to the harmonic trap $x^2/2$, have a quasi-2D profile in $y-z$ plane. 
All parameters are the same as 
in figure \ref{fig2} except the interspecies scattering length 
  $a$(Dy-Er) = $30a_0$. 
The density on contour is 0.001.
}\label{fig7} \end{center}

\end{figure}

Now we consider the shape of binary dipolar soliton with asymmetric traps on 
each of the two components. The easiest way to achieve this is to apply the 
asymmetric trap $x^2/2$ on each of the components. Such a binary soliton is free to move in the $y-z$ plane. Because of the only trap in the $x$ direction 
it is thin in this direction and has a large size in the $y-z$  plane. {The extension along the polarization $z$ direction is the largest as the aligned dipoles along this direction lead to a natural  elongation along this direction for large dipolar interaction, as clearly demonstrated in figures 1 and 3 of 
\cite{cr}, where it is experimentally observed  that  an increase  of the dipolar parameter $a_{dd}/a$ leads to an elongation of the BEC 
along the polarization $z$ direction.  The change in the dipolar parameter was achieved by changing the scattering length $a$ by the Feshbach resonance technique \cite{fesh}.} 
The profiles of the 1000 $^{164}$Dy and the 1000 $^{168}$Er atoms in this binary soliton 
are quite similar and are illustrated in figures \ref{fig7} (a) and (b), respectively, for the $^{164}$Dy and $^{168}$Er components.

\begin{figure}[!t]

\begin{center}
\includegraphics[width=.325\linewidth]{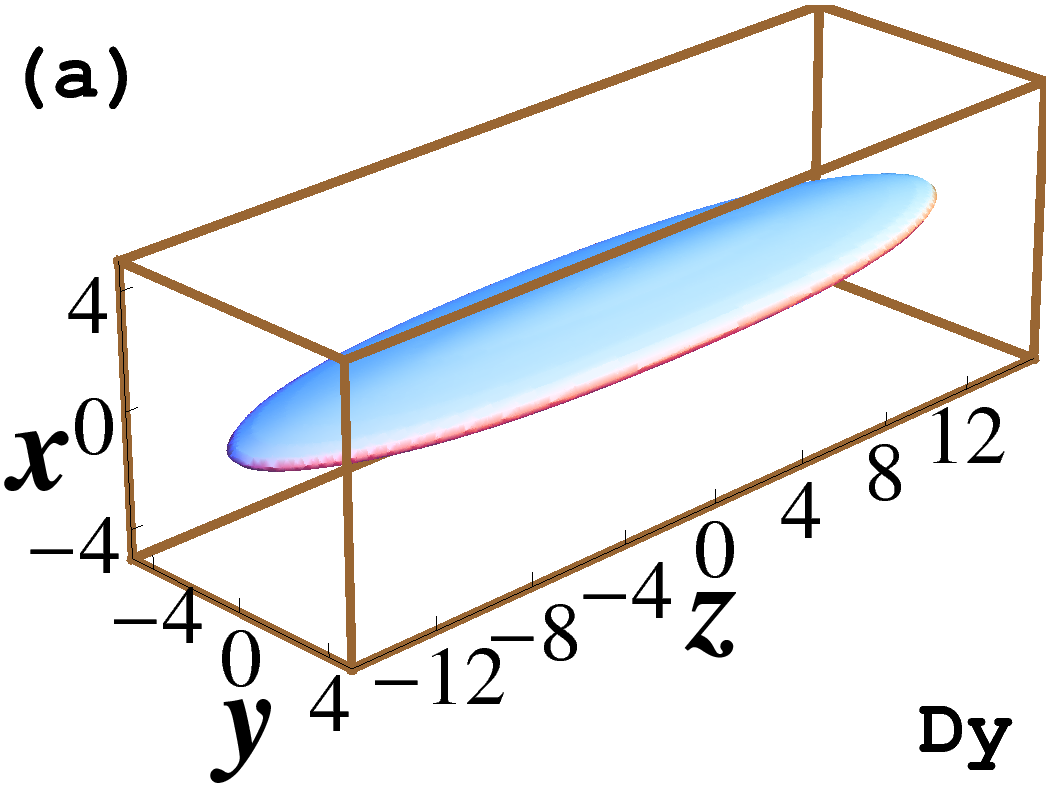}
\includegraphics[width=.325\linewidth]{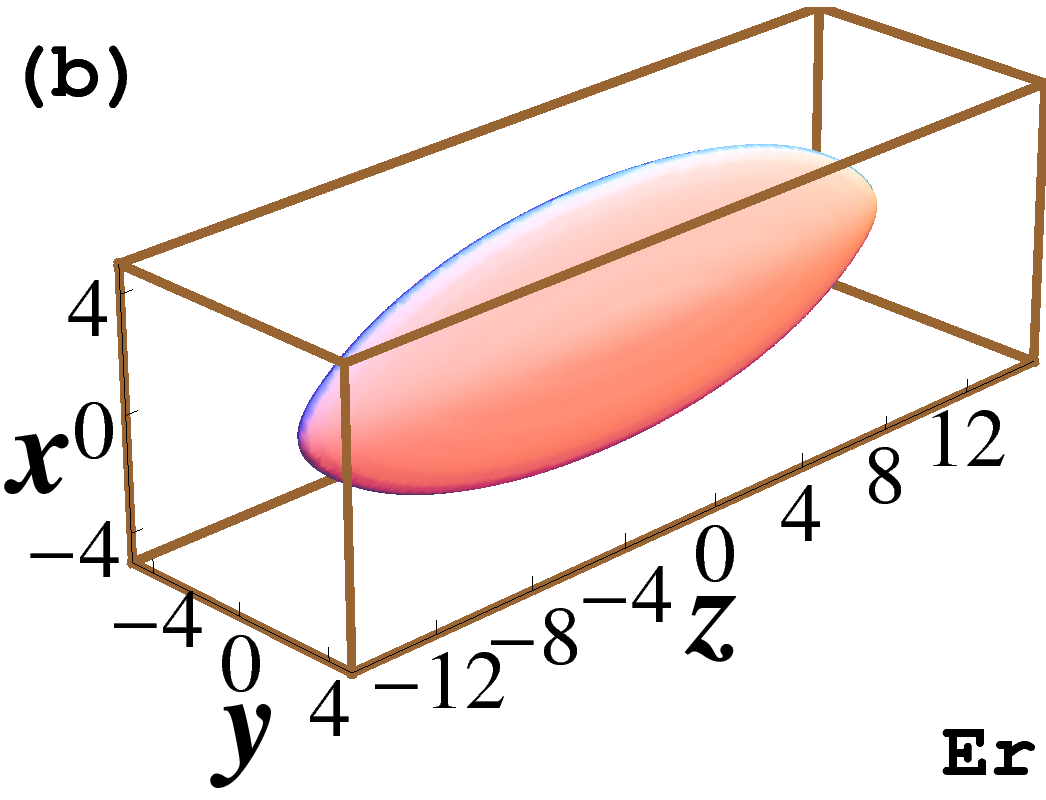}
\includegraphics[width=.325\linewidth]{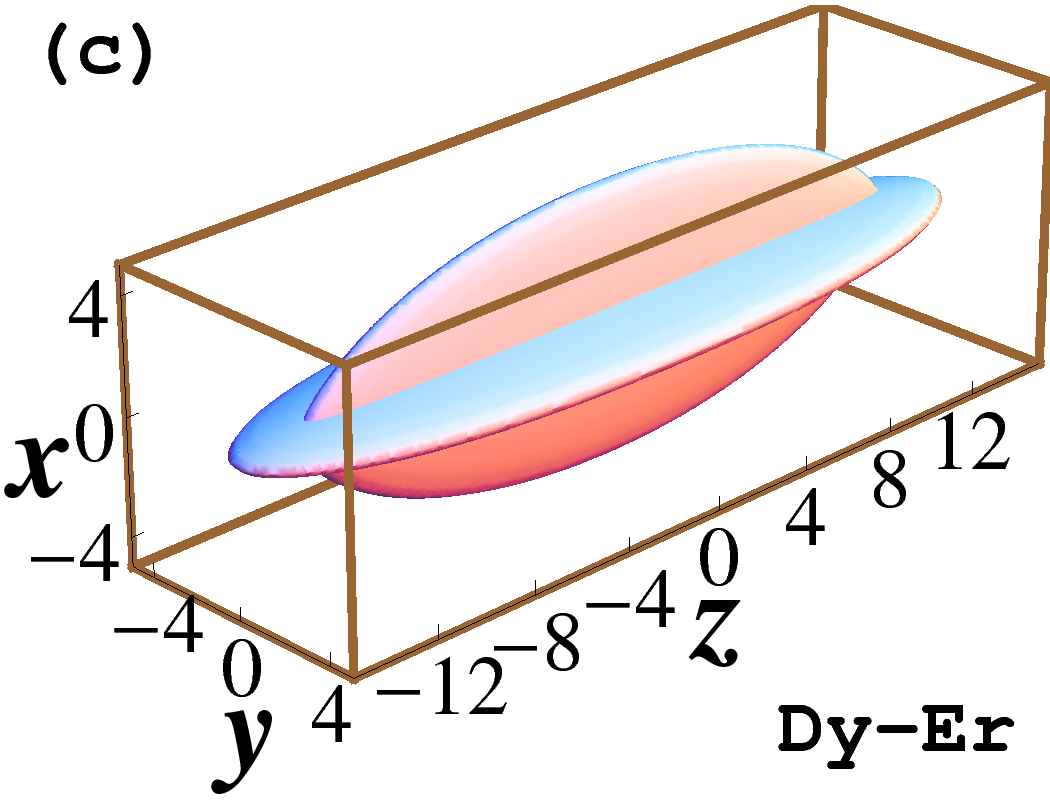}

\caption{ 
 Isodensity plots of numerically obtained 
(a) $^{164}$Dy and (b) 
$^{168}$Er and (c)   $^{164}$Dy-$^{168}$Er
profiles of a binary 
$^{164}$Dy-$^{168}$Er soliton
of 1000 atoms each and 
free to move in the $z$ direction.
 The first component 
(Dy), subject to a harmonic trap $x^2/2$, has a quasi-2D 
profile in $y-z$ plane whereas 
the second component (Er) with the  trap $y^2/2$ has a 
 quasi-2D profile in the  $x-z$ plane.  
All parameters are the same as 
in figure \ref{fig2} except the interspecies scattering length 
  $a$(Dy-Er) = $20a_0$. 
The density on contour is 0.001.
}\label{fig8} \end{center}

\end{figure}

 The binary solitons considered 
so far have a qualitatively similar shape for the two components and hence they are best illustrated in two different plots. 
Next we consider two more types of asymmetric binary solitons which lead to very distinct profiles for the two components. In these cases, in addition to distinct plot for the two components, it is also illustrative to plot the isodensity profiles of the binary soliton.  First we consider an
asymmetric  harmonic trap $x^2/2$ on the first component (Dy) and an asymmetric trap $y^2/2$ on the second component (Er).  The binary soliton is free to move in the $z$ direction with the quasi-2D profile of the 
first component lying in the $y-z$ plane and the quasi-2D profile of the 
second component lying in the $x-z$ plane.  Again due to the aligned dipoles 
in the polarization $z$ direction the binary soliton has the largest spatial 
extension along this direction. The isodensity contour of the two components 
and of the binary soliton are illustrated in figures \ref{fig8} (a), (b), and 
(c), respectively.

\begin{figure}[!t]

\begin{center}
\includegraphics[width=.32\linewidth,clip]{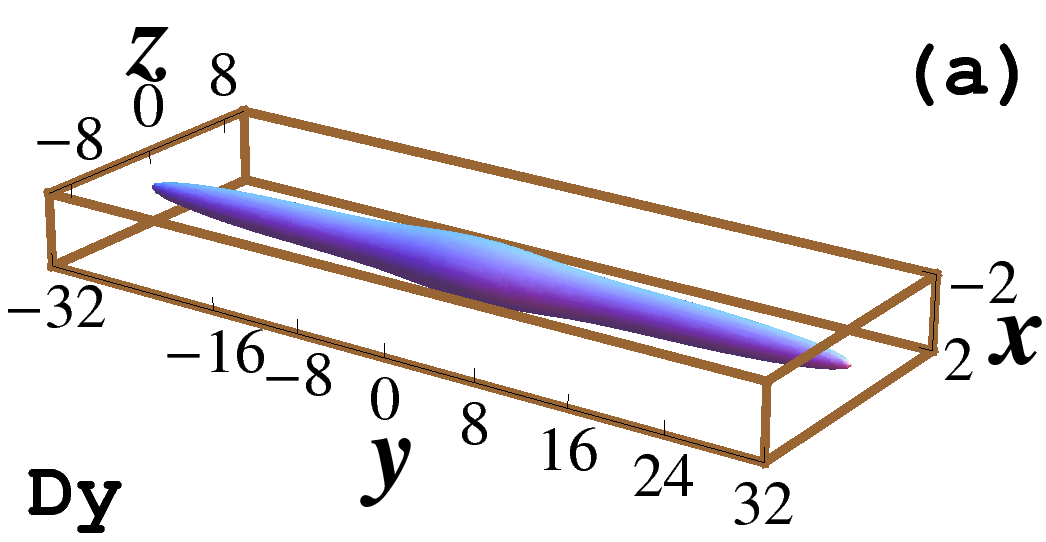}
\includegraphics[width=.32\linewidth,clip]{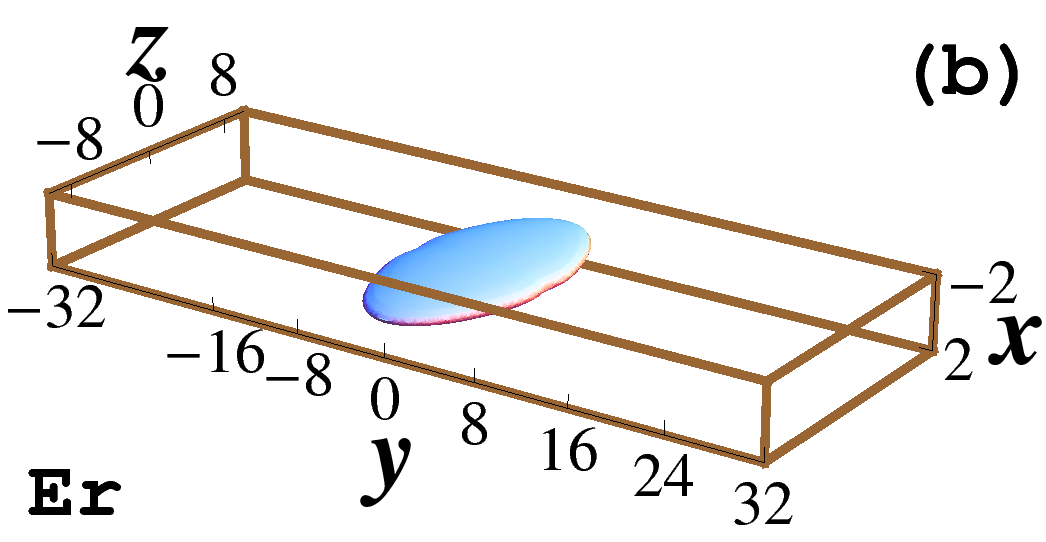}
\includegraphics[width=.32\linewidth,clip]{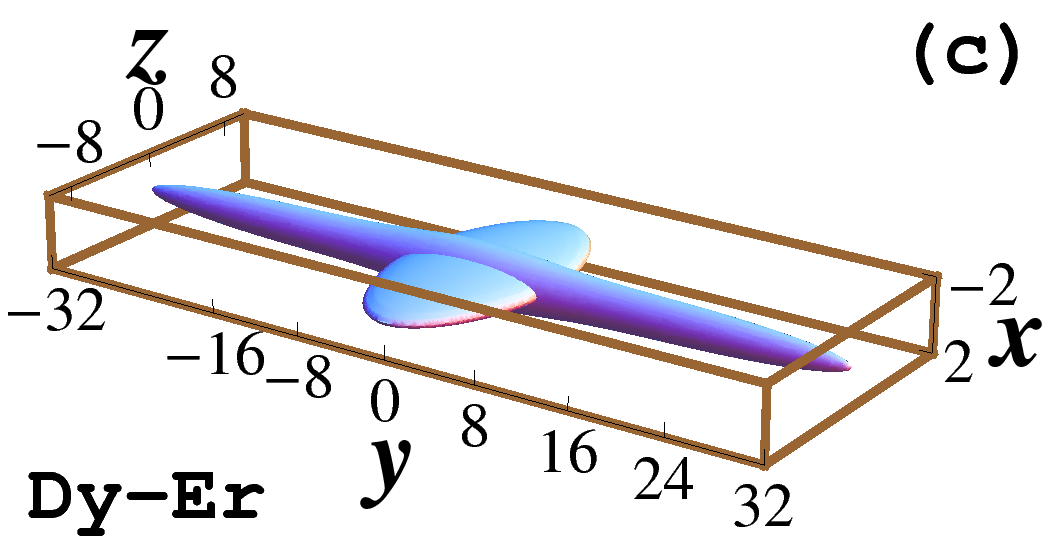}

\caption{  Isodensity plots of numerically obtained 
(a) $^{164}$Dy and (b) 
$^{168}$Er and (c)   $^{164}$Dy-$^{168}$Er
profiles of a binary 
$^{164}$Dy-$^{168}$Er soliton of 200 $^{164}$Dy atoms and 
 1000 $^{168}$Er atoms  
free to move in the $y$ direction.
 The first component 
(Dy), subject to a harmonic trap $(x^2+z^2)/2$, 
has a quasi-1D profile along $y$ direction whereas 
the second component (Er) with the  trap $x^2/2$ has a 
 quasi-2D profile in the  $y-z$ plane.  
All parameters are the same as 
in figure \ref{fig2} except the interspecies scattering length 
  $a$(Dy-Er) = $-40a_0$. 
The density on contour is 0.001.
}\label{fig9} \end{center}
\end{figure}

The binary solitons considered so far have repulsive  interspecies
and intraspecies contact interactions. These solitons are solely confined 
by the dipolar interaction. Now we consider a new class of binary soliton 
stabilized by attractive interspecies contact interaction. In this case the 
first component $^{164}$Dy is subject to the harmonic trap $(x^2+z^2)/2$ 
and the second component $^{168}$Er is subject to the trap $x^2/2$. The interspecies scattering length is set at $a$(Dy-Er)$= -40a_0$. The shapes 
of the two components are quite distinct in this case.
The isodensity profiles of  $^{164}$Dy and $^{168}$Er  are 
shown in figures \ref{fig9} (a) and (b) and that of the binary overlapping 
soliton is illustrated in figure \ref{fig9} (c).

\begin{figure}[!t]

\begin{center}
\includegraphics[width=.49\linewidth,clip]{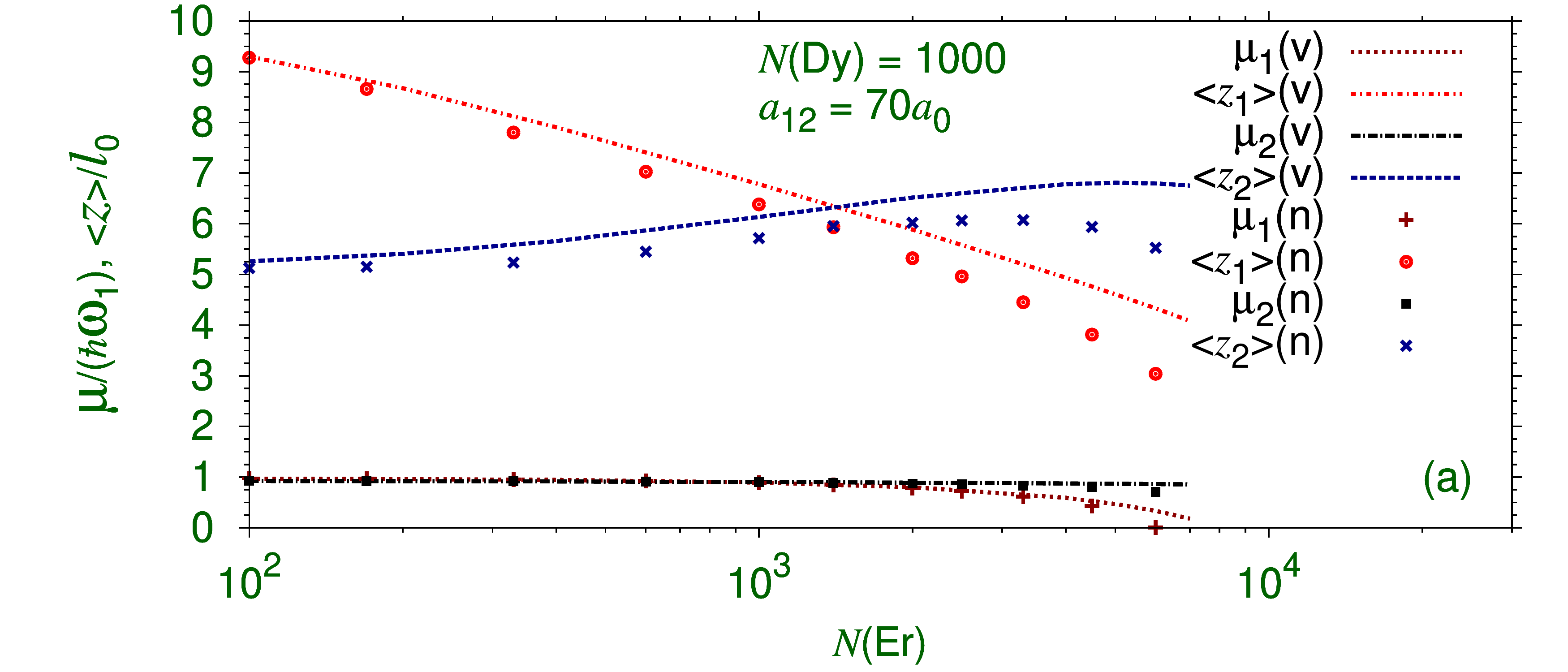}
\includegraphics[width=.49\linewidth,clip]{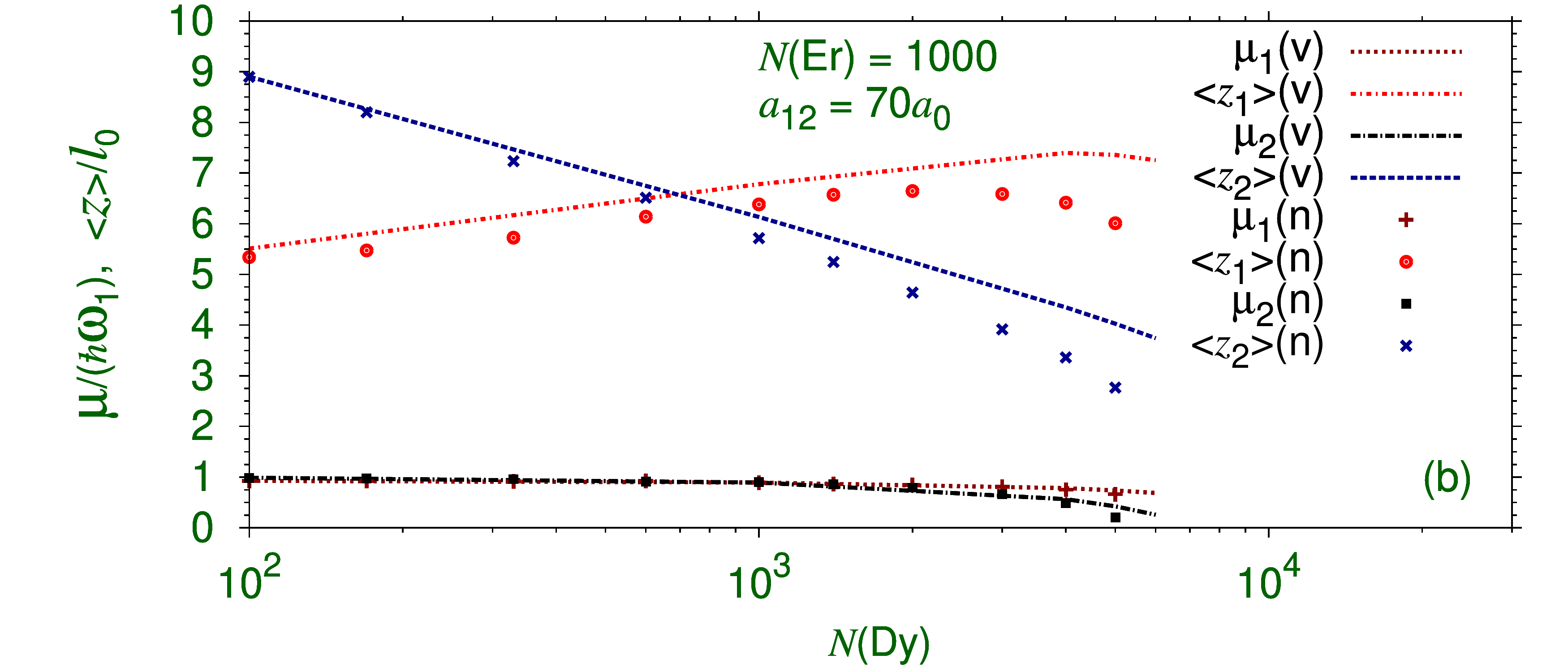}
\caption{  Variational (v) and numerical (n) results
for chemical potential $\mu_i$  and rms size  
 $\langle z_i\rangle $ for the binary $^{164}$Dy-$^{168}$Er
soliton 
versus (a) $N$(Er) for $N$(Dy) $= 1000$, and 
 (b) $N$(Dy) for $N$(Er) $= 1000$, 
with the harmonic trap $\rho^2/2$ on both components.
All parameters are the same as 
in figure \ref{fig2} except the interspecies scattering length 
  $a$(Dy-Er) = $70a_0$. 
    }
\label{fig10}\end{center}

\end{figure}

To test the variational approximations (\ref{eq10}), (\ref{eq11}),
(\ref{eq12}), and (\ref{eq13}), we use them to study the statics and dynamics 
of  the axially-symmetric binary $^{164}$Dy-$^{168}$Er solitons with the radial 
trap $\rho^2/2$ on both components 
and compare 
the results with the numerical solution of  (\ref{eq3}) and (\ref{eq4}). 
We fix the number of either $^{164}$Dy or $^{168}$Er atoms in the binary soliton 
to be 1000 and vary 
the number of the other type of atoms for the fixed interspecies scattering length 
$a_{12}=70a_0$. In this fashion we calculate the chemical potential and root-mean-square (rms) sizes $\langle x\rangle$,   $\langle y\rangle$, and  $\langle z\rangle$,  of the binary solitons.  Because of the radial trap $\rho^2/2$, rms sizes $\langle x\rangle$ and    $\langle y\rangle$ converge 
pretty rapidly and the variational estimates for these sizes agree well with 
the numerical results. On the other hand, the lack of trap in the $z$ direction
makes the convergence of the rms size $\langle z\rangle$ very difficult and the 
variational estimates show larger discrepancies when compared with the 
numerical solution. Hence we present results for the rms size $\langle z\rangle$ and the chemical potential for the binary solitons as obtained from 
variational approximation and numerical solution. In figure \ref{fig10} (a) 
we present these results versus the number of $^{168}$Er atoms $N$(Er) for a fixed number $N$(Dy) = 1000 of   $^{164}$Dy atoms. In figure \ref{fig10} (b) 
we show the same versus the number of $^{164}$Dy atoms $N$(Dy) for a fixed number $N$(Er) = 1000 of   $^{168}$Er atoms. From figures \ref{fig10} we find 
that the agreement between the variational and numerical results is satisfactory. The larger discrepancy occurs for larger number of atoms in the 
binary soliton. For larger number of atoms, the large nonlinear interactions 
in the binary soliton change the profile of the solitons from the assumed 
Gaussian shape in the variational approximation thus leading to a larger 
discrepancy between variational and numerical estimates.

\begin{figure}[!t]

\begin{center}
\includegraphics[width=.49\linewidth]{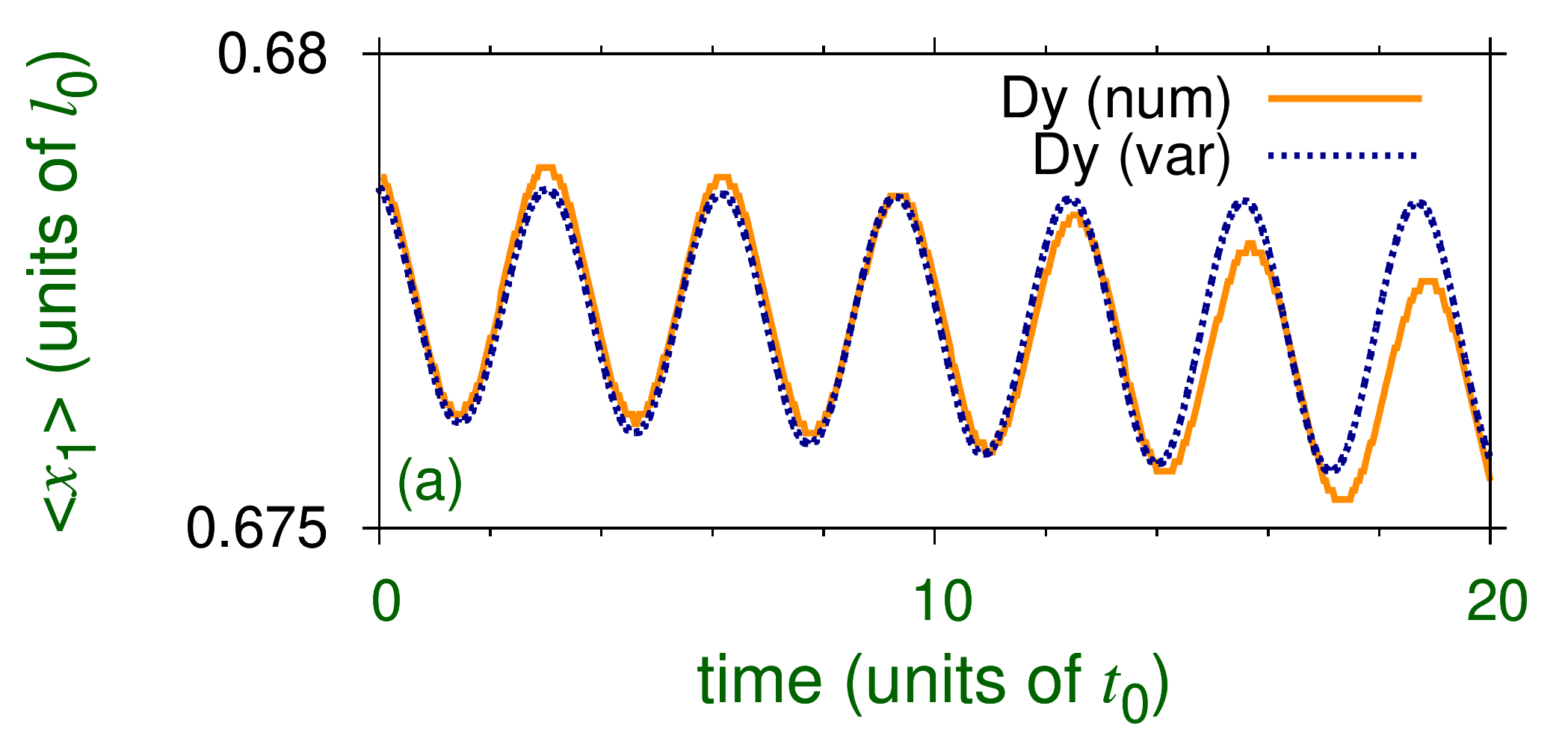}
\includegraphics[width=.49\linewidth]{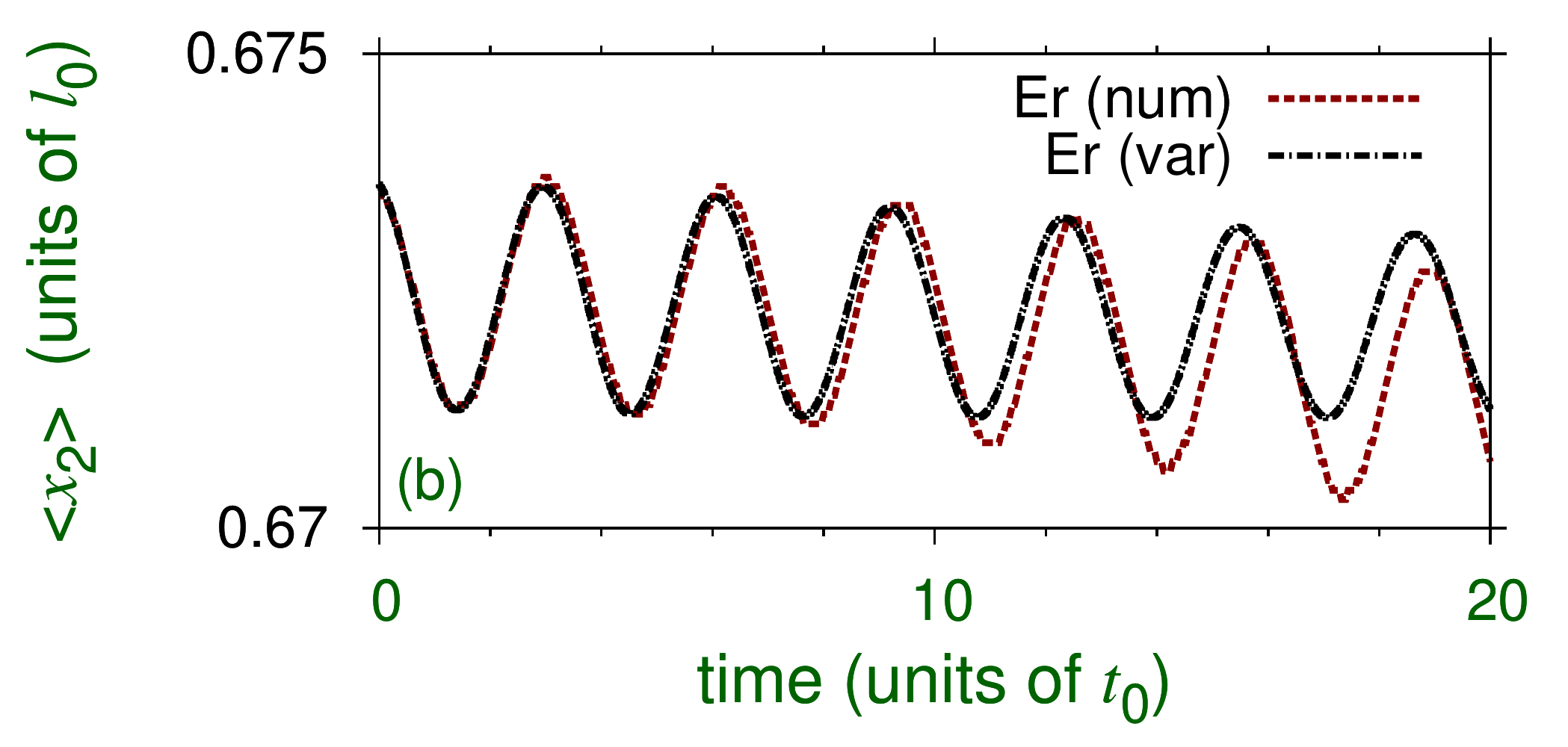}
\caption{  Numerical (num) and variational (var)
time evolutions of rms size $\langle x \rangle $
of (a) $^{164}$Dy and (b) $^{168}$Er components in the 
binary soliton of figure \ref{fig5} upon a small perturbation. }
\label{fig11}\end{center}

\end{figure}

We investigate the dynamics of an axially-symmetric
 binary  $^{164}$Dy-$^{168}$Er soliton %of figure \ref{fig4}
free to move along the axial $z$ 
direction and radially trapped by the potential $\rho^2/2$.
A small oscillation is generated by an infinitesimal perturbation 
in the numerical routine. The oscillation in the radial direction is 
quasi sinusoidal, whereas the oscillation in the axial direction is 
found to be more complicated. This is due to the presence of the harmonic trap 
in the radial direction. In figures \ref{fig11} (a) and (b), to show  
the radial oscillation of the two components of the binary soliton we 
plot the time evolution of the rms size $\langle x \rangle$ of the two components $-$ $^{164}$Dy and $^{168}$Er, respectively.
The same oscillations generated from the variational equations
(\ref{eq10}) $-$ (\ref{eq13}) are also shown in these figures.
Considering the complicated dynamics the agreement between the two results is 
quite satisfactory.

\begin{figure}[!t]

\begin{center}
\includegraphics[width=\linewidth]{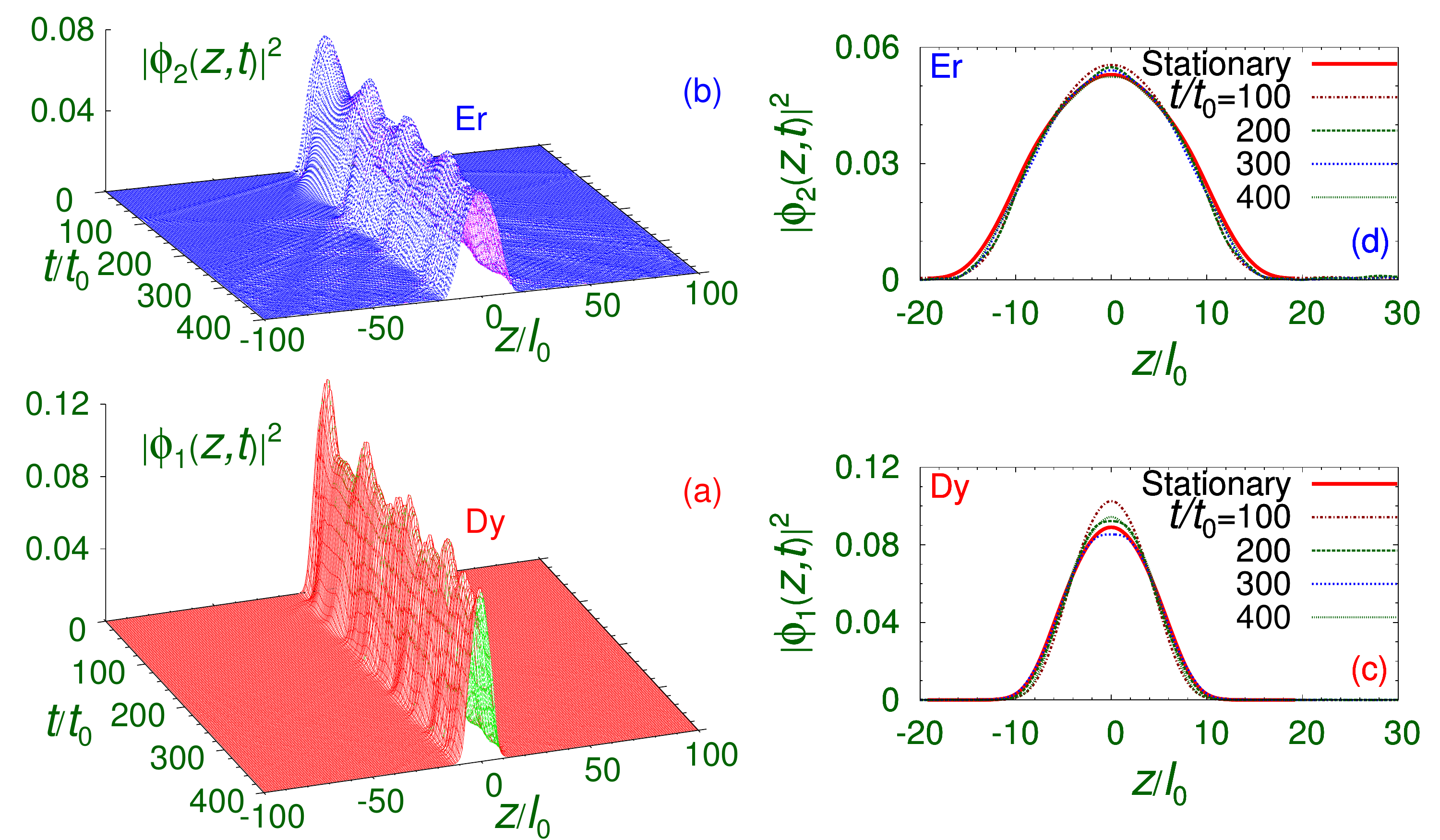}

\caption{ Integrated 1D density $|\phi_i(z; t)|^2 =
\int dx \int dy |\phi_i({\bf r},t)|^2$ 
 of (a) Dy and (b) Er atoms during real-time propagation
when the axial trap of angular frequency $\omega_z = 2 \pi \times 
3$ 
Hz on a quasi-1D axially-symmetric binary  dipolar BEC of 1000 Dy and 5000 Er atoms
is removed linearly for $20 > t/t_0 > 0$ and the resultant soliton
is propagated for $t/t_0>20$. 
The profiles of the 1D density of (c) Dy and (d) Er atoms at times $t/t_0 = 100, 200,300$ and 400
during real-time propagation compared with the stationary 1D density of the binary soliton 
obtained from imaginary-time propagation (Stationary). 
Parameters used in simulation: $a_{dd}($Dy$) =
132.7a_0; a_{dd}($Er$) = 66.6a_0; a_{dd}($Dy-Er$) = 94.0a_0; a($Dy$) =
120a_0; a($Er$) = 60a_0; a($Dy-Er$) = 70a_0; \omega_\rho = 2\pi \times 61$ Hz. }
\label{fig12}\end{center}

\end{figure}

{ As the dipolar solitons are robust and stable, they can be prepared and studied experimentally relatively easily as compared to nondipolar solitons. To obtain a quasi-1D dipolar binary soliton free to move along axial $z$ direction and bound by the harmonic trap $\rho^2/2$ in the $x-y$ plane, first,   we prepare a quasi-1D binary dipolar BEC bound under an weak axial trap: $V({\bf r})=(\rho^2+\lambda^2z^2)/2, \lambda = 0.05$, corresponding to  radial and axial traps with angular frequencies $\omega_\rho = 2\pi \times 61$ Hz and $\omega_z \approx 2\pi \times 3$ Hz.  Then we slowly and linearly remove the axial trap in 20 units of dimensionless time $t/t_0$, when the bound binary dipolar BEC turns to a binary dipolar BEC soliton. To illustrate the viability of this scheme we prepare the axially-symmetric binary dipolar BEC 
wave functions in imaginary-time propagation of Eqs. (\ref{eq3}) and (\ref{eq4}) with $\nu_1=\gamma_1=\nu_2=\gamma_2=1$ and $\lambda_1=\lambda_2 = 0.05$.  The dipolar binary BEC wave functions obtained in imaginary-time simulation is then used in the real-time routine to study the dynamics of the preparation of the dipolar binary soliton. 
During real-time propagation,
from $t/t_0$ = 0 to 20 the axial trap $\lambda^2z^2/2$ is gradually
(linearly) reduced to zero, so that for $t/t_0 > 20$ the  
 axially free quasi-1D binary dipolar soliton emerges. Then we continue the real-time propagation for 
a large interval of time and establish a stable binary soliton at large times. The result of the   simulation is presented in figure \ref{fig12}, where we plot the integrated 1D densities $|\phi_i(z,t)|^2 = \int dx \int dy |\phi({\bf r},t)|^2$ during real-time propagation. We further confirm that the profiles of the binary dipolar soliton oscillated a little during real-time propagation 
around the stationary  soliton profiles obtained from imaginary-time propagation. To this end, 
we show in figures \ref{fig12} (c) and (d) the 1D density of Dy and Er atoms  at times 
$t/t_0 = 100, 200, 300, $ and 400 together with the converged
stationary imaginary-time profiles. 
The stable and robust peaks of the component solitons  confirm the solitonic nature of the binary  BEC. In this fashion all the binary solitons could be realized in laboratory, by including a weak trap along the free propagation direction(s) and eventually realeasing these traps slowly.  A nondipolar BEC soliton stabilized by contact attraction alone 
cannot be prepared in this fashion. }

\section{Summary and Discussion} 
\label{IV}

Using variational approximation and numerical solution of a set of coupled 3D mean-field GP equations, 
we demonstrate the existence
 of a  dipolar binary  $^{164}$Dy-$^{168}$Er  soliton stabilized  by
inter- and intraspecies  
dipolar interactions in the presence of repulsive
 inter- and intraspecies contact interactions.
  The domain of stability of the binary soliton 
 is highlighted in stability phase diagrams of number of $^{164}$Dy and $^{168}$Er 
 atoms and interspecies scattering length
$a_{12}$ for fixed dipolar interactions and intraspecies contact interactions.
We considered distinct spatial shapes  of the two components of the 
binary dipolar soliton, e.g., (a) both with quasi-1D profile along the polarization $z$ direction, (b) both with quasi-2D profile in the $y-z$ plane,
(c) one component  with quasi-1D profile along the $z$ direction and 
the other with quasi-2D profile in the $y-z$ plane, (d) one component  with quasi-2D profile in the $y-z$ plane and the other with quasi-2D profile in the $x-z$ plane,  etc. 
  Results of variational approximation and numerical solution for
the statics (sizes and chemical potentials) and dynamics (breathing oscillation) of the binary  $^{164}$Dy-$^{168}$Er  soliton   with
both components having axially-symmetric quasi-1D profile
are found to be in 
satisfactory agreement with each other.

The solitons considered in this work are stabilized by long-range dipolar 
attraction and short-range contact repulsion. The repulsion in the dipolar 
interaction is equilibrated by a harmonic  trap. Hence unlike normal BEC 
solitons stabilized by short-range contact attraction alone, the present dipolar 
BEC solitons will be more immune to collapse due to short-range repulsion
and can easily accommodate 10000 atoms of the  binary $^{164}$Dy-$^{168}$Er mixture as can be
seen from Figs. \ref{fig2} and \ref{fig4}.
A possible way of preparing these binary dipolar solitons is suggested. 
As these solitons are robust and stable one can first prepare the binary dipolar BECs  with weak 
traps along the mobile directions. Then these weak traps are to be removed slowly, when the binary BECs will turn to binary solitons. The viability of this approach is 
demonstrated  by real-time 
simulation. 
Hence these solitons could be of great experimental interest. With the present experimental techniques, 
such binary dipolar BECs can be observed and the conclusions of the present study verified.

\ack

%\acknowledgments
We thank  
FAPESP  and  CNPq (Brazil)  for partial support.


\section*{References}


\begin{thebibliography}{99}

\bibitem{1}  Strecker K E,  Partridge G B,  Truscott A G  and  Hulet R G
2002  {\it Nature} {\bf 417} 150 


\bibitem{2}  Khaykovich L,  Schreck F,  Ferrari G,  Bourdel T,  Cubizolles J,  Carr L D,  Castin Y  and   Salomon C 2002 
{\it Science} {\bf 256} 1290 
\bibitem{3}  Cornish S L,  Thompson S T  and  Wieman C E 2006 
\PRL  {\bf 96} 170401 


\bibitem{4}
Perez-Garcia V M,  Michinel H  and  Herrero H 1998 \PR A  {\bf 57} 3837



\bibitem{dy} Lu M,  Youn S H and  Lev B L 2010 \PRL
 {\bf 104}
063001 

 McClelland  J J and  Hanssen J L 2006 \PRL {\bf 96} 143005 

  Youn S H,  Lu M W, 
Ray U and  Lev B L 2010 \PR A {\bf 82} 043425 


\bibitem{ExpDy}Lu M,  Burdick N Q,  Youn S H  and  Lev B L 2011 {\it \PRL} {\bf 107} 190401 


\bibitem{ExpEr}
Aikawa K,  Frisch A,  Mark M,   Baier S,  Rietzler A,
 Grimm R and  Ferlaino F 2012 {\it Phys. Rev. Lett.} {\bf 108} 210401



\bibitem{crrev} Lahaye T  {\it et al.} 2009 {\it Rep. Prog. Phys.} {\bf 72} 126401 


\bibitem{cr}  Lahaye T {\it et al.} 2007  {\it Nature} {\bf 448} 672 

 

\bibitem{saddle}  
Stuhler J {\it et al.} 2005  \PRL {\bf 95} 150406 

\bibitem{ExpCr} Goral K,  Rzazewski K and  Pfau T 2000 \PR A {\bf 61} 051601 



\bibitem{52Cr} Koch T {\it et al.} 2008  {\it Nature Phys.} {\bf 4} 218 



Griesmaier A {\it et al.} 2006  \PRL {\bf 97} 250402 


\bibitem{polar} Deiglmayr J,  Grochola A, 
Repp M,  M\"ortlbauer K,  Gl\"uck C,  Lange J,  Dulieu O,  Wester R
and  Weidem\"uller M 2008 {\it Phys. Rev. Lett.} {\bf 101} 133004 

 de Miranda M H G {\it et al.} 2011 {\it Nature Phys.} {\bf 7} 502 


 

\bibitem{1D}
 Young-S L E,  Muruganandam P  and  adhikari S K 2011  \jpb \textbf{44}
101001 



 \bibitem{2D}
 Pedri P and  Santos L 2005
\PRL {\bf 95} 200404  (2005)

 Tikhonenkov I,  Malomed B A  and  Vardi A 2008 \PR A {\bf 78} 043614 

 Tikhonenkov I,  Malomed B A and  Vardi A 2008   \PRL {\bf 100} 090406 

 K\"oberle P,  Zajec D,  Wunner G and  Malomed B A 2012
\PR A {\bf 85} 023630 



\bibitem{ol1D}Adhikari S K and  Muruganandam P 2012
\textit{Phys. Lett. A } \textbf{ 376} 2200 


\bibitem{ol2D}
Adhikari S K and  Muruganandam P 2012
\jpb \textbf{45} 045301 

\bibitem{mfb} Wilson R M,  Ticknor C,  Bohn J L and  Timmermans E 2012
\PR
  A {\bf 86} 033606 


  Saito H,  Kawaguchi Y and  Ueda M 2009 \PRL {\bf 102}
230403 

\bibitem{mfb2} Young-S L E and Adhikari S K 2012
\textit{\PR A }{\bf 86} 063611 

\bibitem{ex}
 Lamporesi G {\it et al.} 2010 \PRL
 {\bf 104} 153202 





\bibitem{th} Young-S L E,  Salasnich L  and   Adhikari S K 2010
 \PR A {\bf 82}  053601 


 
\bibitem{17}
 Yi S and  You L 2001  \PR A {\bf 63} 053607 

 
%
\bibitem{Santos01} Goral K and  Santos L 2002 \PR A. {\bf 66} 023613 


\bibitem{pg}  Perez-Garcia V M,  Michinel H,   Cirac J I,  Lewenstein M  and  Zoller P 1997 \PR A {\bf 56}
1424 

 
\bibitem{CPC}  Muruganandam P and  Adhikari S K  2009 {\it Comput. Phys. Commun.}
{\bf 180}
 1888 

 Vudragovic D,  Vidanovic I,  Balaz A,  Muruganandam P and  Adhikari S K 2012 {\it Comput. Phys. Commun.}
{\bf 183}
 2021 


\bibitem{fesh}
Inouye S {\it et al.} 1998 {\it Nature} {\bf 392} 151 


%\bibitem{collapse}T. Lahaye {\it et al.}, \PRL {\bf 101}, 080401 %(2008).


 

\bibitem{opfesh}
Blatt S,  Nicholson T L,  Bloom B J,  Williams J R,  Thomsen J W,  Julienne P S and  Ye J  2011 \PRL {\bf 107} 073202 



%\bibitem{lattice}R. M. W. van Bijnen, D. H. J. O'Dell, N. G. Parker, and A. M. Martin,
%\PRL {\bf 98}, 150401 (2007);
%R. Kishor Kumar, P. Muruganandam, J. Phys. B {\bf 45},  215301 (2012);
%M. Abad, M. Guilleumas, R. Mayol, M. Pi, and D. M. Jezek,
%\PR A {\bf 79}, 063622 (2009).
  
 
\end{thebibliography}
\end{document}